\documentclass[11pt]{amsart}
\usepackage{amsaddr}
\usepackage{graphicx,amssymb,amsmath,amsthm}

\usepackage[utf8]{inputenc}
\usepackage{epstopdf}
\usepackage{bm}
\usepackage[numbers]{natbib}
\usepackage{color}
\usepackage{array}

\usepackage{graphicx,amsmath,amsfonts}
\usepackage{epstopdf}

\usepackage{lipsum}
\usepackage{kantlipsum} 

\calclayout

\usepackage{amsfonts}
\usepackage{graphicx}
\usepackage{epstopdf}
\usepackage{algorithmic}
\usepackage{bm} 

\usepackage{amsopn}

\newcommand{\R}{\mathbb{R}}

\newcommand{ \E}{\mathbb{E}}
\newcommand{ \V}{\mathbb{V}}

\newcommand{\ha}{\frac{1}{2}}
\newcommand{\pdif}[2]{\frac{\partial #1}{\partial #2}}

\newcommand{\Ord}{\mathcal{O}}

\newcommand{\deta}{d\hat \eta}
\newcommand{\ddeta}{d\tilde\eta}

\newcommand{\re}[1]{\textrm{Re}\left( #1\right)}
\newcommand{\imag}[1]{\textrm{Im}\left( #1\right)}

\newcommand{\sechsq}[1]{\operatorname{sech}^2\left(#1\right)}

\newtheorem*{Definition}{Definition}
\usepackage[mathscr]{euscript}

\usepackage{amsfonts}

\title[Collective coordinate framework for stochastic partial differential equations]{A collective coordinate framework to study the dynamics of travelling waves in stochastic partial differential equations}
\author{Madeleine C. Cartwright and Georg A. Gottwald}
\address{School of Mathematics and Statistics, University of Sydney, NSW 2006, Australia}
\email[M. C. Cartwright and G. A. Gottwald]{M.Cartwright@maths.usyd.edu.au {\rmfamily and} georg.gottwald@sydney.edu.au}

\begin{document}

\begin{abstract}
We propose a formal framework based on collective coordinates to reduce infinite-dimensional stochastic partial differential equations (SPDEs) with symmetry to a set of finite-dimensional stochastic differential equations which describe the shape of the solution and the dynamics along the symmetry group. We study SPDEs arising in population dynamics with multiplicative noise and additive symmetry breaking noise. The collective coordinate approach provides a remarkably good quantitative description of the shape of the travelling front as well as its diffusive behaviour, which would otherwise only be available through costly computational experiments. We corroborate our analytical results with numerical simulations of the full SPDE.
\end{abstract}



\maketitle


\section{Introduction}
Stochastic partial differential equations (SPDEs) are a standard part of a scientist's toolbox to study the effect of fluctuations and to include mesoscopic effects in natural and engineered systems. Applications range from genetics, epidemic outbreaks, and population dynamics to chemical engineering \cite{RiordanEtAl95,GarciaOjalvo,Kotelenez,Kuehn13,EtheridgeEtAl14}. The inclusion of noise can have profound impact on the dynamics of the propagation of coherent structures \cite{SchimanskyGeierZuelicke91,VanSarloos03,Panja04}.\\ 
We treat here SPDEs where the deterministic part exhibits symmetry; in particular, we study SPDEs with translational symmetry supporting travelling waves. We set out here to reduce the complexity of such infinite-dimensional systems and propose a formal framework to reduce the dynamics of travelling fronts in SPDEs to a set of finite-dimensional stochastic differential equations (SDEs). For travelling waves such reductions were studied in the limit of small noise amplitudes for the Fisher-KPP equation and bistable systems by mathematicians and physicists \cite{MuellerSowers95,ConlonEtAl05,MuellerEtAl11,Funaki95,MikhailovEtAl83,SchimanskyGeierZuelicke91}. Rather than employing perturbative expansions to study small noise perturbations \cite{MikhailovEtAl83,SchimanskyGeierZuelicke91,RoccoEtAl02,MendezEtAl11} or judiciously chosen moment closure schemes \cite{Hallatschek11}, we adopt here the perspective of decomposing the dynamics into the dynamics on the symmetry group and the dynamics orthogonal to it, established for deterministic partial differential equations (PDEs) with symmetry (see for example \cite{GolubitskyStewart}). This symmetry perspective of spatially extended systems has proved successful in studying and classifying pattern formation \cite{Barkley94,GottwaldMelbourne13}, constructing efficient numerical methods for equivariant systems \cite{BeynThuemmler04,HermannGottwald10,FoulkesBiktashev10}, and in studying Hamiltonian systems such as planetary dynamics \cite{RobertsWulffLamb02} and observed spectra of ${\rm{CO}}_2$ molecules \cite{CushmanEtAl04}.\\ Within the symmetry perspective PDEs are cast into a {\em{skew product}} of ordinary differential equations whereby the dynamics on the symmetry group is driven by the so-called shape dynamics.  In the case of a travelling wave with underlying translational symmetry this amounts to the following standard situation: the shape dynamics is an equilibrium solution in the frame of reference moving with constant wave speed, and the dynamics on the translation group orbit describes the linear drift of the reference frame in physical space. In PDEs with translational symmetry in a $d$-dimensional physical space, the skew product reduces to
\begin{align}
\dot v &= g(v)
\label{e.x}
\\
\dot \phi &= h(v),
\label{e.phiskew}
\end{align}
where $\phi\in\R^d$ represents the translation variables. If the shape dynamics \eqref{e.x} consists of an equilibrium $v(t)\equiv v_0$ we obtain $\phi(t)=c t$ with $c=h(v_0)$. This includes the case of a travelling wave moving witorganizh constant speed $c$ mentioned above.\\

This paper is concerned with employing this framework to SPDEs with weak noise. The main idea of this work is that the effect of the noise will be controlled in the shape dynamics which is dominated by strong contraction to the travelling wave solution; along the neutral direction on the group orbit, however, the noise is unconstrained, leading to diffusive behaviour of the location of the front. Employing this idea we will express the solution of the SPDE in terms of judiciously chosen collective coordinates, describing the shape and the group dynamics, respectively, the dynamics of which will be described by SDEs. This reduction of an infinite-dimensional SPDE to a system of finite-dimensional SDEs describes the dynamics of travelling fronts remarkably well as we will demonstrate in numerical simulations. As an explicit example we study the bistable Nagumo equation \cite{NagumoEtAl62,Murray} subjected to multiplicative noise as well as spatially localized additive noise. We show how the collective coordinate approach quantitatively describes the speed of the front propagation and the shape of the front. We find that the addition of noise leads to a slowing down and a steepening of the front in the case of multiplicative noise. Further, we consider the collective coordinate approach in situations where the translational symmetry is broken and study cases where the noise is confined to a bounded interval in space. The collective coordinate approach can be applied inside the noisy spatial region with again remarkably good quantitative skill in describing the dynamics of front propagation.\\

The paper is organized as follows. In Section~\ref{s.models} we introduce the stochastic partial differential equations. We propose in Section~\ref{s.cc} the framework of stochastic collective coordinates, and apply it to the bistable equation with multiplicative noise in Section~\ref{s.bistable_mult} and with additive noise in Section~\ref{s.bistable_add}. We present numerical results illustrating the ability of our approach to capture the effect of multiplicative noise and of localized additive noise on the dynamics. We conclude in Section~\ref{s.discussion} with a discussion and an outlook.


\section{Models}
\label{s.models} 
We study here SPDEs for a single component $u(x,t)$ in one spatial dimension of the form
\begin{align}
\partial_t u = D \partial_{xx}u + f(u) +  \dot \eta(u,x,t),
\label{e.SPDE0}
\end{align}
where the noise has zero mean $ \E\,[ \eta(x,t)]   =  0$. Here the expectation is taken with respect to realizations of the driving noise process. As an explicit example, frequently considered in population dynamics, we consider a bistable reaction term
\begin{align}
f(u) = u(1-u)(u-b)\, ,
\label{e.bistableF}
\end{align}
with $b>0$, modelling the Allee effect when at low densities individual fitness increases with density \cite{MendezEtAl11,Kuehn13}. The resulting SPDE is known as the bistable equation or, in the context of action potential propagation on nerve fibres, as the Nagumo equation \cite{NagumoEtAl62,Murray}. The Nagumo equation supports in the noise-free case travelling waves with stable asymptotic states $u=0$ and $u=1$. We consider additive and multiplicative noise. As multiplicative noise we consider noise which vanishes at the asymptotic states $u=0$ and $u=1$. This prevents nucleation phenomena outside of the front interface. A biologically motivated example of such a noise is a fluctuating Allee threshold $b+\dot B(t)$. In this case the noise is given as 
\begin{align}
\eta(x,t) = \sigma u(1-u)\, B(t) ,
\label{e.noisebistable}
\end{align}
with one-dimensional Brownian motion $B(t)$ \cite{ArmeroEtAl96}. Note that the noise respects the underlying translational symmetry. As additive noise we consider white in time noise with spatial correlations prescribed by a covariance operator, $Q$, with kernel 
${\mathcal{C}}(x,x^\prime)=\exp(-|x-x^\prime |/\ell)$
with finite trace. Then there exists a complete orthonormal basis $\{\varphi_k\}_{k=1}^\infty$ with $Q\varphi_k=\lambda_k\varphi_k$, and we can construct a $Q$-Wiener process
\begin{align}
\eta(x,t) = \sigma s(x)\,  \sum_{k=1}^\infty \sqrt{\lambda_k}\varphi_k(x) B_k(t)
\label{e.QWiener}
\end{align} 
with independent one-dimensional Brownian motions $B_k(t)$ satisfying $ \E\,[\eta] =0$ and \linebreak $ \E\,[ \dot \eta(x,t)\,\dot \eta(x^\prime,t^\prime)] = \sigma^2 s(x)s(x^\prime) \, Q\, \delta(t,t^\prime)$  \cite{DaPratoZabczyk,PrevotRoeckner}. We further assume that the noise is localized in space in a region $x\in [-\tfrac{1}{2}L_{\rm{noise}},\tfrac{1}{2}L_{\rm{noise}}]$, thereby breaking the translational symmetry, which we model by
\begin{align}
s(x) = \frac{1}{2}(1+\tanh(\kappa(x-\tfrac{1}{2}L_{\rm{noise}}))-\tanh(\kappa(x+\tfrac{1}{2}L_{\rm{noise}})))
\label{e.noiselocal}
\end{align} 
with $\kappa \gg 1$.


\section{Stochastic collective coordinates}
\label{s.cc}
Initially developed for conservative deterministic nonlinear wave dynamics \cite{McLaughlinScott78,Scott}, the theory of collective coordinates has since been extended to dissipative deterministic  systems such as reaction diffusion systems \cite{GottwaldKramer04,MenonGottwald05,MenonGottwald07,MenonGottwald09,CoxGottwald06} and recently to phase oscillators \cite{Gottwald15,Gottwald17,HancockGottwald18}. Here we apply the method to the stochastic partial differential equations introduced in the previous section.\\ 

Consider a parabolic semilinear SPDE of the form
\begin{align}
\partial_t u(x,t) = D\Delta u + f(u) + \dot\eta(u,x,t),
\label{e.SPDE}
\end{align}
with noise $\eta(u,x,t)=\sum_{k=1}^M g_k(u,x)\,B_{k}(t)$ with one-dimensional Brownian motion $B_{k}(t)$ and $x\in \mathcal{D}$. $M$ can be finite or infinite. We assume that the solution can be approximated by some ansatz function $\hat u(x,t; \bf{p})$ for some time-dependent, so called collective coordinates ${\bf{p}}\in\R^n$. The particular functional form of the ansatz function has to be judiciously chosen to capture the character of the solution of the SPDEs and can, for example, be inferred from numerical simulations of the SPDE. The dynamics of the infinite-dimensional SPDE is then encoded in the temporal evolution of the finite-dimensional collective coordinates ${\bf{p}}(t)$, which we write in a general form as
\begin{align}
d{\bf{p}} = {\bf{a}}_{\bf{p}}({\bf{p}}) \, dt + \sigma_{\rm{p}}({\bf{p}}) \, d{\bf{W}}(t),
\label{e.cc_ansatz_gen}
\end{align}
where ${\bf{W}}(t)$ is $m$-dimensional Brownian motion. Here the subscripts in the drift term ${\bf{a}}_{\bf{p}}$ and in the $n\times m$ diffusion matrix $\sigma_{\rm{p}}({\bf{p}})$ refer to the collective coordinates, i.e. ${\bf{a}}_{p_{j}}$ denotes the drift term for the collective coordinate $p_j$ and $\sum_k \sigma_{\rm{p_{j}} {k}}({\bf{p}})d{W}_k(t)$ denotes the diffusive terms for the collective coordinate $p_j$. The aim is to find expressions for the drift term ${\bf{a}}_{\bf{p}}$ and the diffusion matrix $\sigma_{\rm{p}}({\bf{p}})$.\\

The ansatz function $\hat u(x,t)$ does not satisfy the SPDE (\ref{e.SPDE}) and inserting it into the SPDE defines a stochastic process via
\begin{align*}
d \mathcal{E}
= d \hat u - \left[D\Delta \hat u + f(\hat u)\right] dt - g_k(\hat u)\,  dB_{k}(t)
\end{align*}
which quantifies the error made by restricting the solution space to the ansatz function (\ref{e.cc_ansatz_gen}) spanned by the collective coordinates. We remark that our definition of the error is different from the definition of the error as the (weighted) $L^2$ norm of the difference $u(x,t)-\hat u(x,t)$ which is typically used in the literature on stochastic travelling fronts. Employing It\^o's formula on the smooth $\hat u$ we write 
\begin{align*}
d\mathcal{E}(x,t;{\bf{p}})
= \pdif{\hat u}{p_j} d p_j + \frac{1}{2} dp_l \pdif{^2\hat u}{{p_l} {\partial{p_j}}} dp_j 
- \left[D\Delta \hat u + f(\hat u)\right] \, dt -g_k(\hat u)\,  dB_{k}(t) ,
\end{align*}
where we used Einstein's summation convention. Substituting (\ref{e.cc_ansatz_gen}) and collecting only terms up to order $dt$ we obtain, using the independence of the Brownian motion, 
\begin{align*}
d\mathcal{E}(x,t;{\bf{p}}) 
= \left[\pdif{\hat u}{p_j} a_{p_j} + \frac{1}{2} \sigma_{{\rm{p_l}} k}\pdif{^2\hat u}{{p_l} {\partial{p_j}}} \sigma_{{\rm{p_{j}}}  {k}} -D\Delta \hat u - f(\hat u)\right] dt 
+ \left[\pdif{\hat u}{p_j} \sigma_{{\rm{p_{j}}} {k}}\, dW_{k}(t)-g_k(\hat u)\,  dB_{k}(t)\right] .
\end{align*}
To maximize the degree to which the collective coordinates approximate solutions of the SPDE, we require that the stochastic process defined by $d\mathcal{E}$ does not project onto the subspace spanned by the collective coordinates. We therefore require that the error $d\mathcal{E}$ is orthogonal to the tangent space of the solution manifold which is spanned by $\pdif{\hat u}{{p_i}}$ , $i=1,\cdots,n$. Projecting the error eliminates the spatial dependency and yields a system of $n$ algebraic equations for the drift and diffusion coefficients. We separate these conditions in terms corresponding to drift and to diffusion, 
respectively. The $n$ drift contributions are given by 
\begin{align}
\langle \frac{\partial \hat u}{\partial p_i} \frac{\partial \hat u}{\partial p_j}\rangle  a_{p_j} 
+ \frac{1}{2} \langle \frac{\partial \hat u}{\partial p_i} \pdif{^2\hat u}{{p_l} {\partial{p_j}}}\rangle  \sigma_{\rm{{p_l}} k }\sigma_{\rm{p_{j}} {k}} 
-D\langle  \frac{\partial u}{\partial p_i} \Delta \hat u \rangle  
- \langle \frac{\partial u}{\partial p_i} f(\hat u)\rangle
&= 0
\label{e.cc_drift_gen}
\end{align}
for $i=1,\cdots,n$ and the $n$ diffusion contributions, which balance the Brownian motion of the SPDE with the Brownian motion of the collective coordinate system, are given by
\begin{align}
\langle \frac{\partial \hat u}{\partial p_i} \pdif{\hat u}{p_j} \rangle  \sigma_{\rm{p_{j}} {k}}\, dW_{k}(t) =  \langle \frac{\partial \hat u}{\partial p_i} g_k(\hat u) \rangle \,  dB_{k}(t)
\label{e.cc_noise_gen}
\end{align}
for $i=1,\cdots,n$. Here the angular brackets denote integration over the spatial domain as in $\langle A(x)\rangle = \int_\mathcal{D} A(x)dx$. For $m=M$ and $W_{k}(t)=B_{k}(t)$ for all $k=1,\cdots,M$ we can achieve pathwise matching provided
\begin{align}
\sigma_{\rm{p_{j}} {k}} = \frac{\langle \frac{\partial \hat u}{\partial p_i} g_k(\hat u) \rangle}{\langle \frac{\partial \hat u}{\partial p_i} \pdif{\hat u}{p_j} \rangle}.
\label{e.sigma_path0}
\end{align}
The $n$ equations for the drift coefficients $a_{p_i}$ (\ref{e.cc_drift_gen}) and the $nm=nM$ diffusion coefficients (\ref{e.sigma_path0})  determine the drift and diffusion coefficients in the evolution equation for the collective coordinates (\ref{e.cc_ansatz_gen}). Note that the noise in the system of SDEs for the collective coordinates is typically multiplicative since the diffusion coefficients depend on the collective coordinates ${\bf{p}}$.\\ 

In the case when $M$ is infinite, one may want to consider finite $m$ to reduce the complexity of the system of SDEs (\ref{e.cc_ansatz_gen}) for the collective coordinates. Rather than achieving pathwise approximation of the solutions with $W_{k}(t) = B_{k}(t)$ for all $k$, one may require instead that the statistical behaviour of the solution is reproduced. The left-hand side of (\ref{e.cc_noise_gen}) defines the vector-valued stochastic process $P_{i}(t) = \int_0^t S_{ik} dW_k(t)$ and the right-hand side the stochastic process $N_{i}(t) = \int_0^t G_{ik} dB_k(t)$ with $S_{ik} = \langle \frac{\partial \hat u}{\partial p_i} \pdif{\hat u}{p_j} \rangle  \sigma_{\rm{p_{j}{k}}}$ and $G_{ik}=\langle \frac{\partial \hat u}{\partial p_i} g_k(\hat u) \rangle$. Both processes are mean-zero. Moreover, for 
\begin{align*}
S S^T = G G^T
\end{align*}
with $\E\int |S|^2dt<\infty$ and $\E\int |G|^2dt<\infty$, the respective covariances of the two processes are also equal with $\E[ PP^T] =\E[NN^T]$. The condition $S S^T = G G^T$ then allows for a unique solution (up to the sign) of the diffusion matrix $\sigma_{\rm{p}}$ of the collective coordinates as a matrix square root. Again, together with the $n$ equations for the drift coefficients $a_{p_i}$ (\ref{e.cc_drift_gen}) this determines the drift and diffusion coefficients in the evolution equation for the collective coordinates (\ref{e.cc_ansatz_gen}).\\

If the SPDE is invariant under the action of some symmetry group, we expect deterministic behaviour of the collective coordinates associated with the shape dynamics for sufficiently small noise amplitudes; this is due to the shape dynamics being strongly contracting. On the other hand, the noise is unrestricted along the neutral direction of the group dynamics resulting in diffusive behaviour of the collective coordinates associated with the symmetry group.\\ We remark that the collective coordinate approach is different from the variational approach adopted in \cite{InglisMacLaurin16} or those aimed at solving only for the wave speed \cite{SchimanskyGeierZuelicke91,BenguriaDepassier96,MendezEtAl11} or the interface location \cite{AntonopoulouEtAl12}.\\ 

In the following we apply this general framework to the bistable stochastic partial differential equation introduced in the previous section. 


\section{Collective coordinate approach for the bistable Nagumo equation with multiplicative noise}
\label{s.bistable_mult}
\begin{figure}[htbp]
\centering
\includegraphics[width = 0.6\columnwidth]{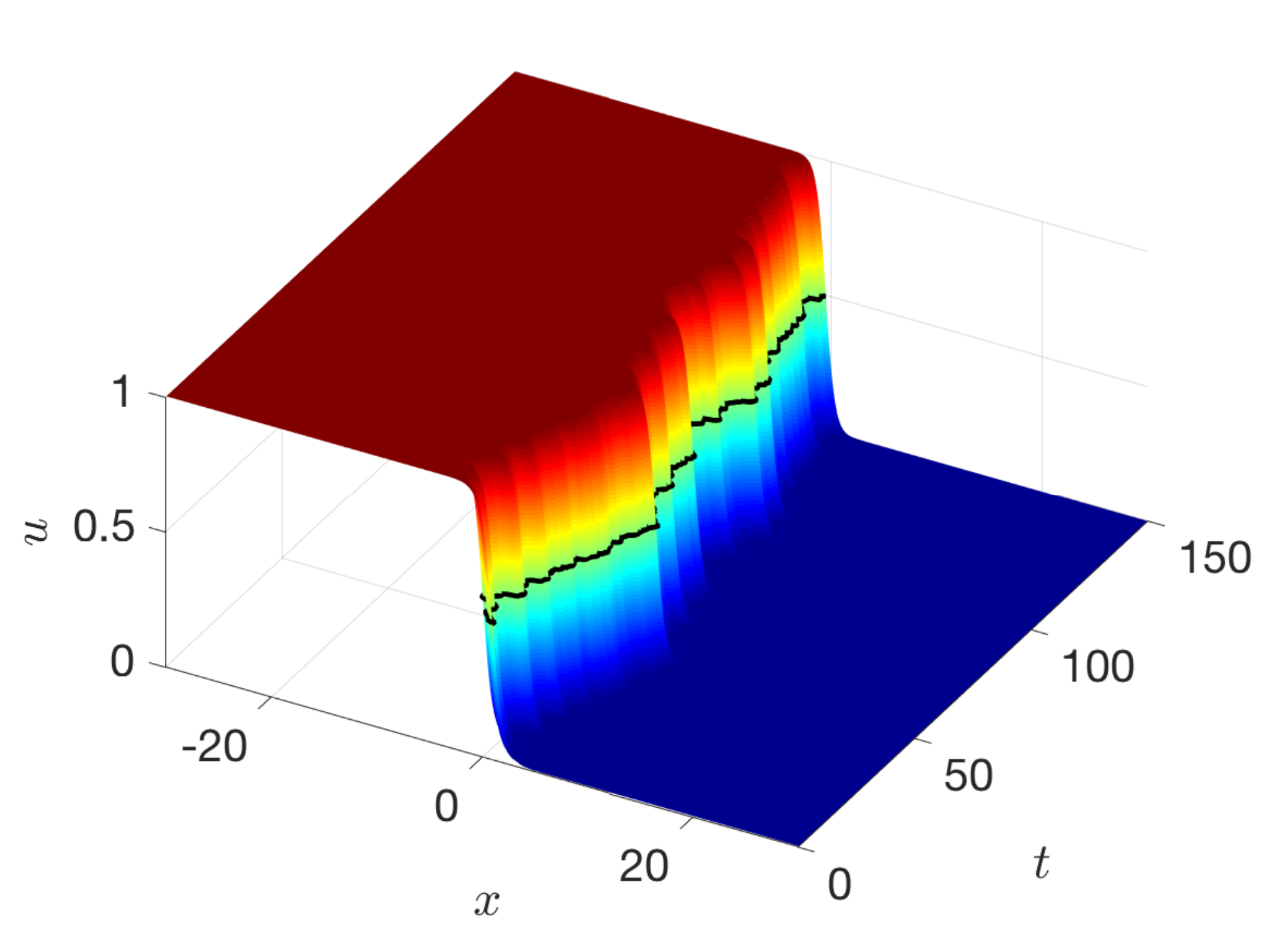}\\
\includegraphics[width = 0.35\columnwidth]{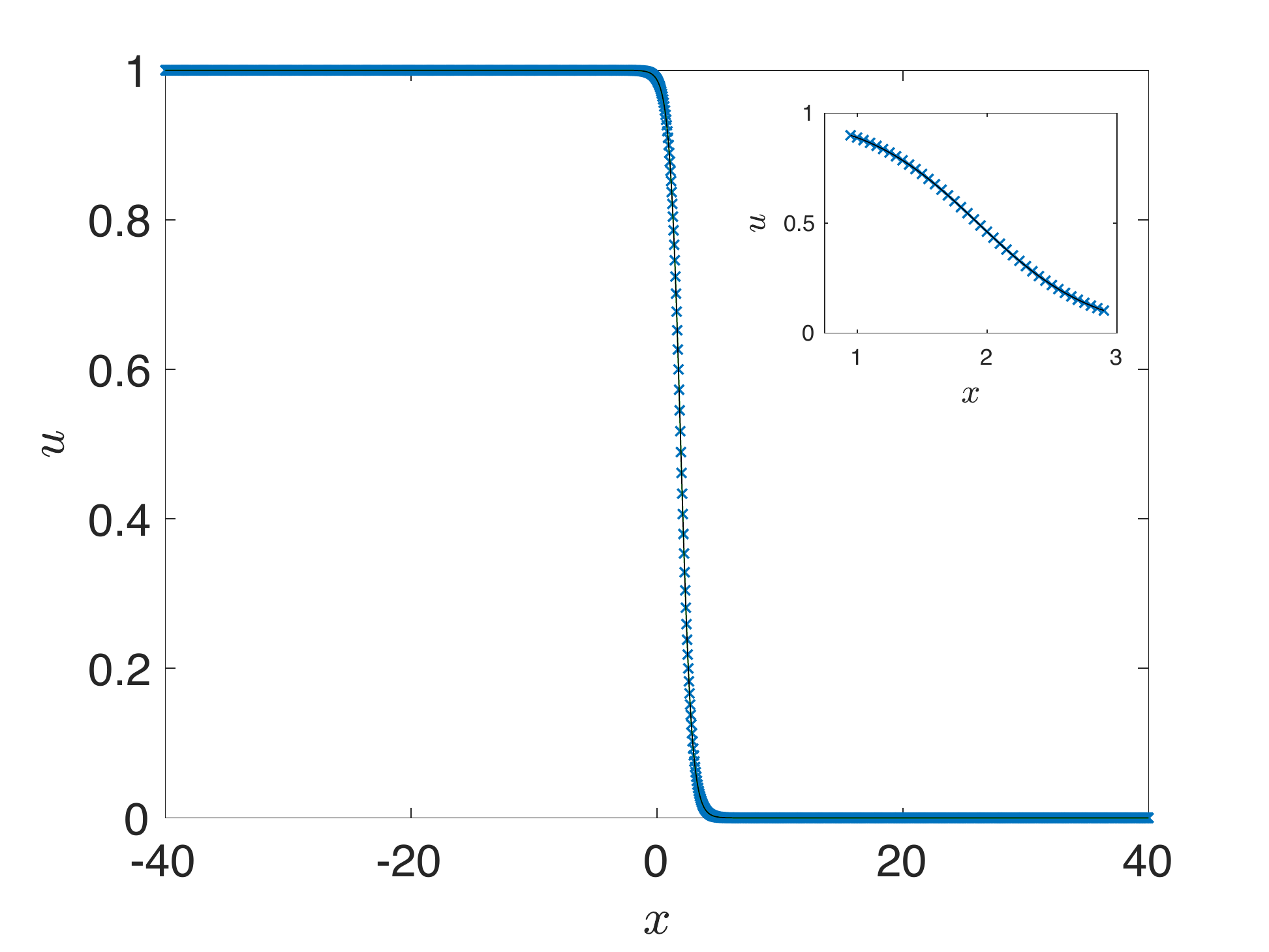}
\includegraphics[width = 0.35\columnwidth]{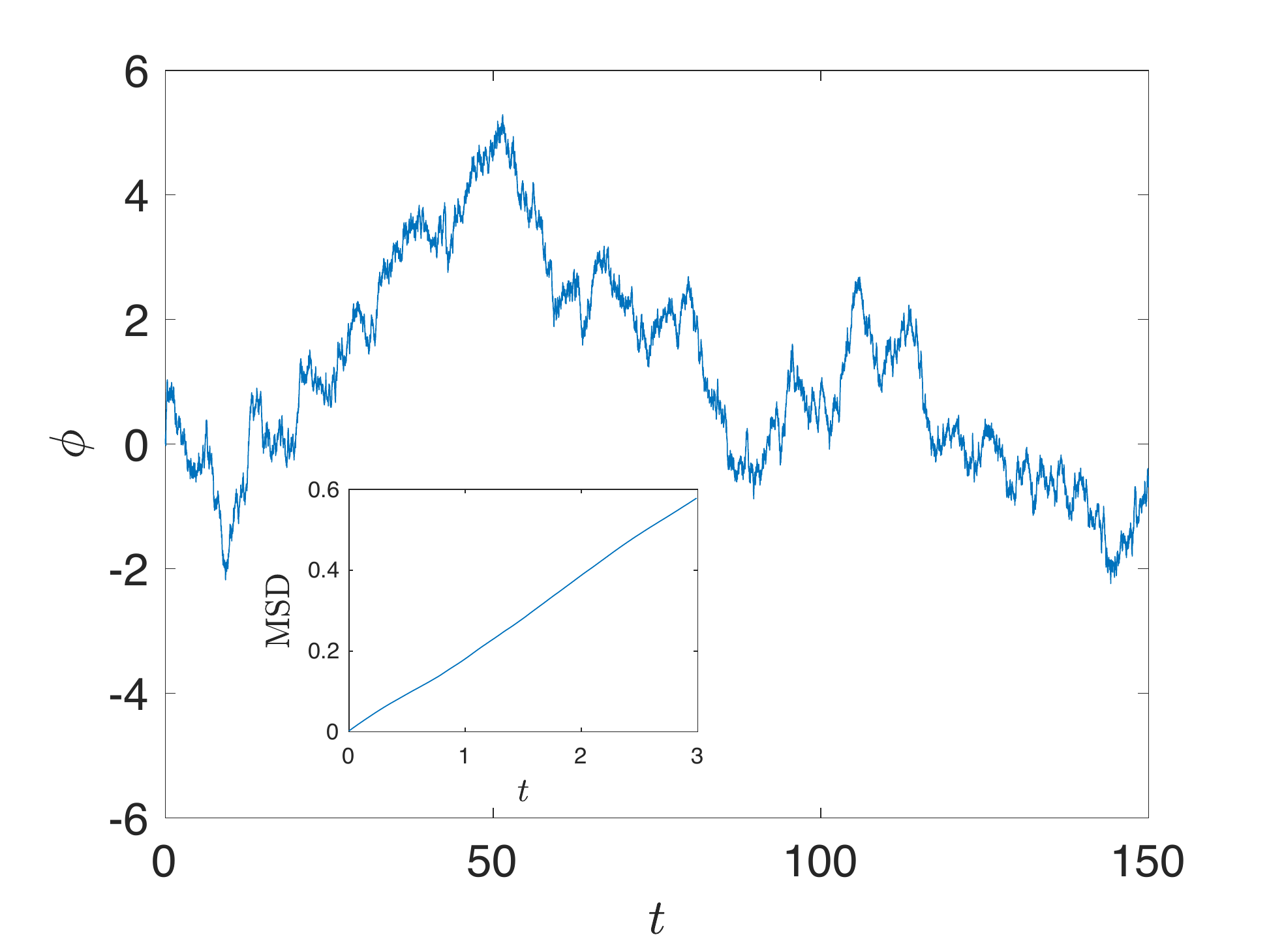}
\caption{Plot of a travelling front solution for the bistable equation (\ref{e.bistable}) with $D=0.2$ and $b=0.5$ driven by multiplicative noise with $\sigma=1$. Top: Space-time plot; the black line denotes the contour $u=0.5$. Bottom Left: Solution for one realization of the noise at fixed time $t>10$; the inset shows a zoom into the interface where the crosses depict the solution of the SPDE and the continuous line is a fit to a $\tanh$-profile. Bottom Right: Time series of the location of the interface $\phi(t)$ defined as $\phi=\xi$ such that $u(\xi)=0.5$ together with an inset showing the mean square displacement demonstrating the diffusive character.}
\label{f.TWlog}
\end{figure}

Consider the bistable SPDE with multiplicative noise defined by (\ref{e.noisebistable})
\begin{align}
\partial_t u = D \partial_{xx}u + u(1-u)(u-b) + u(1-u) \dot B(t) .
\label{e.bistable}
\end{align}
In order to find a dimension reduced description of the bistable Nagumo SPDE (\ref{e.bistable}) we make the following ansatz $u(x,t) = \hat u(x;w,\phi)$ with 
\begin{align}
\hat u(x,t) = \frac{1}{2}\left(1-\tanh(w(t)(x-\phi(t))) \right) ,
\label{e.cc}
\end{align}
which defines the collective coordinates ${\bf{p}}=(w(t),\phi(t))$. In the deterministic case this is an exact solution with $w=w_0 = 1/\sqrt{8D}$ and $\phi = c_0\, t$ with $c_0=\sqrt{D/2}(1-2b)$. We remark that it was shown in \cite{Stannat12,Stannat14,KrugerStannat17,HamsterHupkes17,HamsterHupkes18} that such front solutions are stable in the bistable SPDE (\ref{e.bistable}). The inverse front width $w(t)$ constitutes the shape dynamics whereas the location of the front interface $\phi(t)$ denotes the dynamics on the group orbit. Figure~\ref{f.TWlog} shows a snapshot of a travelling front for the bistable equation with multiplicative noise (\ref{e.bistable}) for one realization of the noise. The travelling front is smooth in space and is well fitted by a $\tanh$-profile supporting the assumption that the solution is close to an ansatz function (\ref{e.cc}) in the presence of noise. The actual shape of the front is independent of the realization of the noise and remains fixed in time. The location of the front interface, on the other hand, depends on the noise realization. The location of the front interface $\phi$ (numerically determined as the unique location $\phi=\xi$ such that the fitted smooth $\tanh$-profile has $\hat u(\xi)=0.5$) exhibits diffusive behaviour in time. This supports the notion that the noise is able to diffuse along the neutral direction of the group orbit, but does little to the strongly contracting shape dynamics. The framework of collective coordinates will allow us to reduce this infinite-dimensional model to a finite dimensional model for the front interface width $1/w$ and the spatial location of the interface $\phi$.\\

The collective coordinates encode the dynamics and are propagated via the system of SDEs 
\begin{align}
d w & = a_w(w,\phi)\, dt + \sigma_{w}(w,\phi)\, dW
\label{e.w}
\\
d\phi &=  a_\phi(w,\phi)\, dt + \sigma_{\phi }(w,\phi)\, dW,
\label{e.phi}
\end{align}
with one-dimensional Brownian motions $W$ (cf. (\ref{e.cc_ansatz_gen})). 
 To determine the unknown drift and diffusion coefficients of the evolution equations (\ref{e.w})--(\ref{e.phi}) for the collective coordinates we require that the error made by restricting the solutions of the SPDE to be described by the ansatz function (\ref{e.cc}) is minimized. The stochastic process $\mathcal{E}$, which quantifies this error, is defined by substituting the ansatz (\ref{e.cc}) into the SPDE, employing It\^o's formula for the smooth $\hat u$ to evaluate time-derivatives. For the bistable equation (\ref{e.bistable}) with multiplicative noise we obtain
\begin{align*}
d\mathcal{E} 
= &\left[ \pdif{u}{w} a_w+\pdif{u}{\phi} a_\phi + \frac{1}{2}\pdif{^2u}{w^2}\sigma_{w}^2 
+ \frac{1}{2}\pdif{^2u}{\phi^2} \sigma_{\phi}^2 
+ \pdif{^2u}{w\partial \phi}\sigma_{w}\sigma_{\phi}\right]\, dt
\nonumber \\
& +(\pdif{u}{w}\sigma_{w}+\pdif{u}{\phi}\sigma_{\phi})\, dW 
-\left[Du_{xx} + f(u)\right] \, dt - d\, \eta(x,t) ,
\end{align*}
where we omit for ease of notation the hat to denote the ansatz function (\ref{e.cc}), and where $f(u)$ is given by (\ref{e.bistableF}) for the bistable SPDE and $d\eta(x,t)$ is the multiplicative noise defined in (\ref{e.noisebistable}). To maximize the degree to which the collective coordinates approximate the solution of the SPDE we set, as outlined in Section~\ref{s.cc}, the projection of the stochastic differential of the error $d\mathcal{E}$  onto the tangent space of the solution manifold (\ref{e.cc}) ---  spanned here by $\partial u/\partial w$ and $\partial u/\partial \phi$ --- to zero. Projecting the error onto $\partial u/\partial w$ and $\partial u/\partial \phi$ yields drift contributions given by
\begin{align}
&\langle \left(\pdif{u}{w}\right)^2\rangle a_w 
+\langle \pdif{u}{w}\pdif{u}{\phi}\rangle a_\phi 
+ \frac{1}{2} \langle \pdif{u}{w} \pdif{^2u}{w^2}\rangle \sigma_{w}^2
+ \langle \pdif{u}{w} \pdif{^2u}{w\partial \phi}\rangle \sigma_{w}\sigma_{\phi}
+ \frac{1}{2} \langle \pdif{u}{w} \pdif{^2u}{\phi^2}\rangle \sigma_{\phi}^2
\nonumber \\ 
&\hphantom{aaaaaaaaaaaaaaaaaaaaaaaaaaaaaaaaaaaaaaa}
= \langle \pdif{u}{w}\left(Du_{xx}+u(1-u)(u-b)\right)\rangle
\label{e.proj_w_1}
\end{align}
and
\begin{align}
&\langle \pdif{u}{\phi}\pdif{u}{w} \rangle a_w 
+\langle \left(\pdif{u}{\phi}\right)^2\rangle a_\phi 
+ \frac{1}{2} \langle \pdif{u}{\phi} \pdif{^2u}{w^2}\rangle \sigma_{w}^2
+ \langle \pdif{u}{\phi} \pdif{^2u}{w\partial \phi}\rangle \sigma_{w}\sigma_{\phi }
+ \frac{1}{2} \langle \pdif{u}{\phi} \pdif{^2u}{\phi^2}\rangle \sigma_{\phi}^2
\nonumber \\ 
&\hphantom{aaaaaaaaaaaaaaaaaaaaaaaaaaaaaaaaaaaaaaa}
= \langle \pdif{u}{\phi}\left(Du_{xx}+u(1-u)(u-b)\right)\rangle \, ,
\label{e.proj_phi_1}
\end{align}
as well as diffusion contributions given by
\begin{align}
&\left(\langle \left(\pdif{u}{w}\right)^2\rangle \sigma_{w} + \langle \pdif{u}{w}\pdif{u}{\phi}\rangle \sigma_{\phi}\right)\, dW 
=
\sigma \langle \pdif{u}{w} u(1-u)\rangle\, dB
\label{e.proj_w_2}
\end{align}
and
\begin{align}
&\left(\langle \pdif{u}{\phi}\pdif{u}{w}\rangle \sigma_{w} + \langle \left(\pdif{u}{\phi}\right)^2\rangle \sigma_{\phi }\right)\, dW 
=
\sigma \langle \pdif{u}{\phi} u(1-u)\rangle\, dB \, .
\label{e.proj_phi_2}
\end{align}
The integrals can be explicitly calculated for the $\tanh$-ansatz function (\ref{e.cc}) (see \ref{app.1}). In particular the diffusion contributions (\ref{e.proj_w_2})--(\ref{e.proj_phi_2}) become
\begin{align}
\sigma_{w}\, dW &= 0\\
\sigma_{\phi}\,dW &= \frac{\sigma}{2w}\, dB,
\end{align}
which is solved by $\sigma_{w}=0$ (i.e. no noise in the equation of the shape parameter $w$ (cf. (\ref{e.w})), and by  setting $\sigma_{\phi}=\sigma/2w$. This solution allows for pathwise approximation with $dW=dB$ in the case when the stochastic forcing of the underlying SPDE (\ref{e.bistable}) is known, or for the matching of the statistics of the stochastic forcing of the SPDE for unknown forcing of the SPDE with $dW \neq dB$.  

Evaluating the projections (\ref{e.proj_w_1})--(\ref{e.proj_phi_1}) determines the drift coefficients, and the evolution equation for the collective coordinates becomes the following skew-product system
\begin{align}
dw &= -\frac{3}{4}\frac{w}{w_0^2(\pi^2-6)}\left(w^2-(1+\sigma^2)w_0^2\right)\, dt
\label{e.SDE_bistable_mult_w}\\
d\phi &=  c_0\frac{w_0}{w}\, dt + \frac{\sigma}{2 w}\, dW(t),
\label{e.SDE_bistable_mult}
\end{align}
with $w_0=1/\sqrt{8D}$ and $c_0 = \sqrt{D/2}(1-2b)=(1-2b)/(4 w_0)$ being the stationary inverse width and constant front velocity of the deterministic bistable partial differential equation, and with Brownian motion $W(t)$. The noise only enters the equation for the group variable and does not affect the shape variable $w$, consistent with the symmetry perspective of strongly contracting dynamics of the shape and neutral direction along the group orbit. The equation for the inverse width $w$ is deterministic and supports a stable steady state 
\begin{align}
\bar w = \sqrt{1+\sigma^2}\,w_0 .
\label{e.bistable_mult_w}
\end{align}
The inclusion of a fluctuating Allee threshold leads to a sharper interface $1/\bar w<1/w_0$ when compared to the deterministic travelling wave. This, maybe counter-intuitive, decrease in the interface width in the presence of noise can be explained by considering that fluctuations will push the front interface to the asymptotic values $u=0$ and $u=1$ with a high probability near the edges of the interface leading to a decreased width. After some transient allowing the shape parameter to acquire its equilibrium value $w=\bar w$, the front interface propagates according to (\ref{e.SDE_bistable_mult}) subject to a mean drift with speed
\begin{align}
\bar c_0 = \frac{c_0}{\sqrt{1+\sigma^2}} ,
\label{e.bistable_mult_c0}
\end{align}
with superimposed fluctuations with constant variance $\sigma^2/4\bar w^2$. The inclusion of noise hence leads to a slowing down of the front when compared to the deterministic speed $c_0>\bar c_0$.\\

We remark that, upon expanding in $\sigma$ and letting $t\to \infty$, we recover the result obtained in \cite{SchimanskyGeierZuelicke91} at order $\mathcal{O}(\sigma)$ with $\bar w=w_0$ and $d\phi = c_0 dt +\sigma/(2 \bar w)\,dW(t)$. Our approach captures higher order effects and allows for the study of the temporal evolution.\\ Furthermore, our results confirm previous detailed numerical results performed in \cite{LordThuemmler12} where it was found that the wave becomes steeper with increasing noise and that the speed decreases linearly with $\sigma^2$. This is consistent with our analytical results, which at $\mathcal{O}(\sigma)$ reduce to $\bar w=(1+\sigma^2/2) w_0$ and $d\phi = c_0(1-\sigma^2/2) dt +\sigma/(2 w_0)\, dW(t)$.

\subsection{Numerical results}
We now present numerical results demonstrating that the stochastic collective coordinate approach is capable of quantitatively describing the diffusive behaviour of travelling fronts. We present results for $D=0.2$, $b=0.1$ and $\sigma = 0.75$. We solve the SPDE (\ref{e.bistable}) with multiplicative noise using an Euler-Maruyama scheme whereby the diffusion operator is treated implicitly to allow for larger time steps. 
We choose as the domain $x\in [-60,60]$ with domain length $L=120$ using Dirichlet boundary conditions, and employ a temporal discretization step of $\Delta t=0.01$ and a spatial discretization with grid spacing $\Delta x = 0.05$. We perform simulations with a total length of $T=5000$ time units. The width of the travelling front and the location of the interface $\phi(t)$ where the front assumes the value $u\equiv 0.5$ are determined via nonlinear least square regression using the Matlab function {\em{lsqnonlin}} \cite{Matlab} for the ansatz function (\ref{e.cc}). We have checked that linear interpolation yields similar results when used to determine the location of the interface $\phi(t)$.\\

\begin{figure}[htbp]
\centering
\includegraphics[width = 0.35\columnwidth]{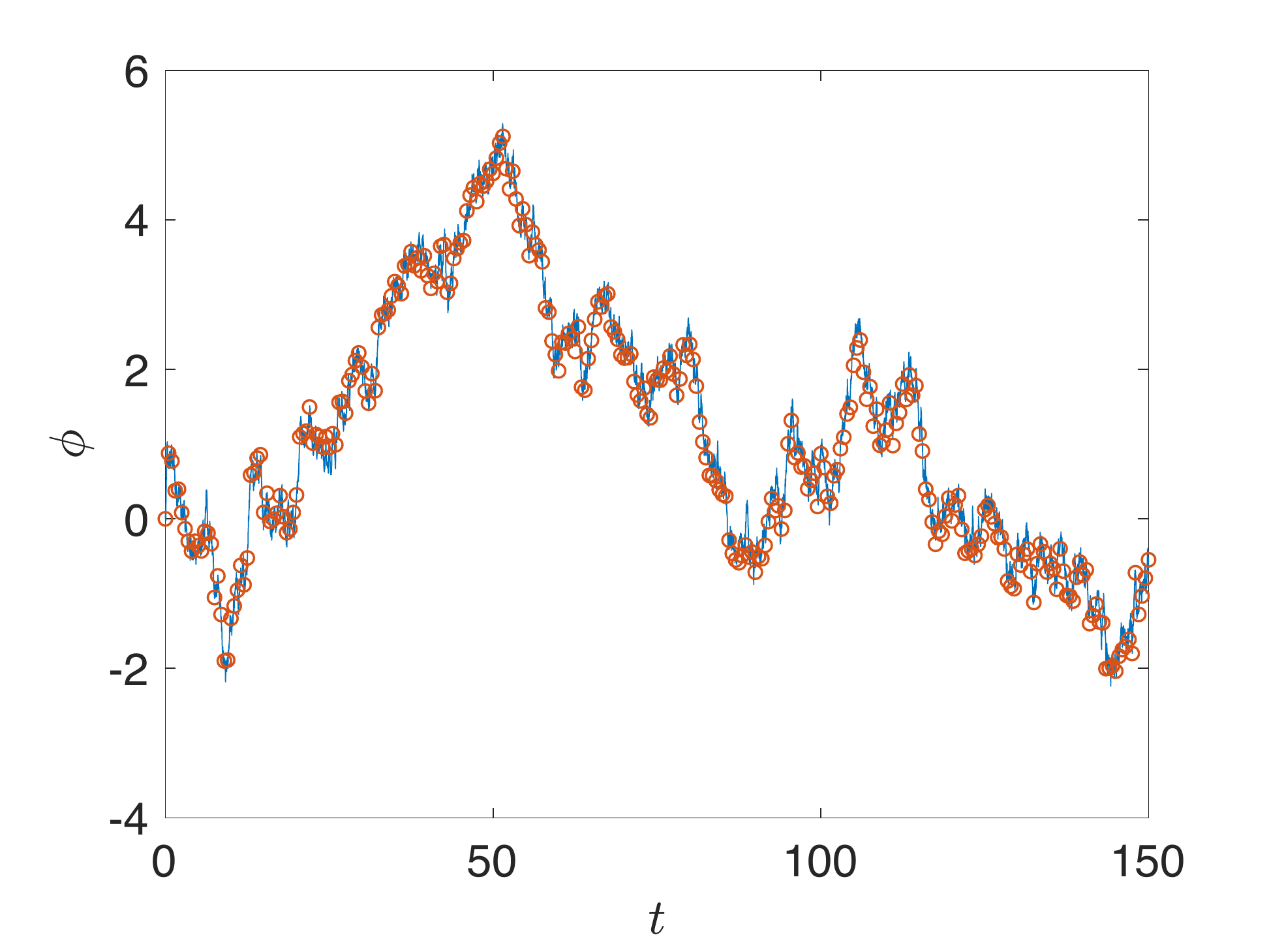}
\caption{Time series of the location of the interface $\phi(t)$ for a travelling front solution for the bistable equation (\ref{e.bistable}) with $D=0.2$ and $b=0.5$ driven by multiplicative noise with $\sigma=1$. Shown are $\phi(t)$ of the solution of the bistable SPDE  (\ref{e.bistable}) defined as $\phi=\xi$ such that the fitted $\hat u(\xi)=0.5$ (blue continuous line) for one realization of the stochastic forcing $B(t)$ as well as the solution of the collective coordinate equation for the phase $\phi(t)$ given by (\ref{e.SDE_bistable_mult}) with $W(t)=B(t)$ (red circles; evaluated at every $50$ time steps to allow for better comparison).} 
\label{f.bistable_mult_pathwise}
\end{figure}

Figure~\ref{f.bistable_mult_pathwise} shows a comparison of the phase of a travelling front of a simulation of the bistable SPDE (\ref{e.bistable}) (defined as $\phi=\xi$ such that the fitted $\hat u(\xi)=0.5$) and its collective coordinate approximation provided by the system of SDEs (\ref{e.SDE_bistable_mult_w})--(\ref{e.SDE_bistable_mult}) when we use the same stochastic forcing for both systems with $W(t)=B(t)$, demonstrating the effectiveness of our approximation to provide pathwise approximation.\\

\begin{figure}[htbp]
\centering
\includegraphics[width = 0.35\columnwidth]{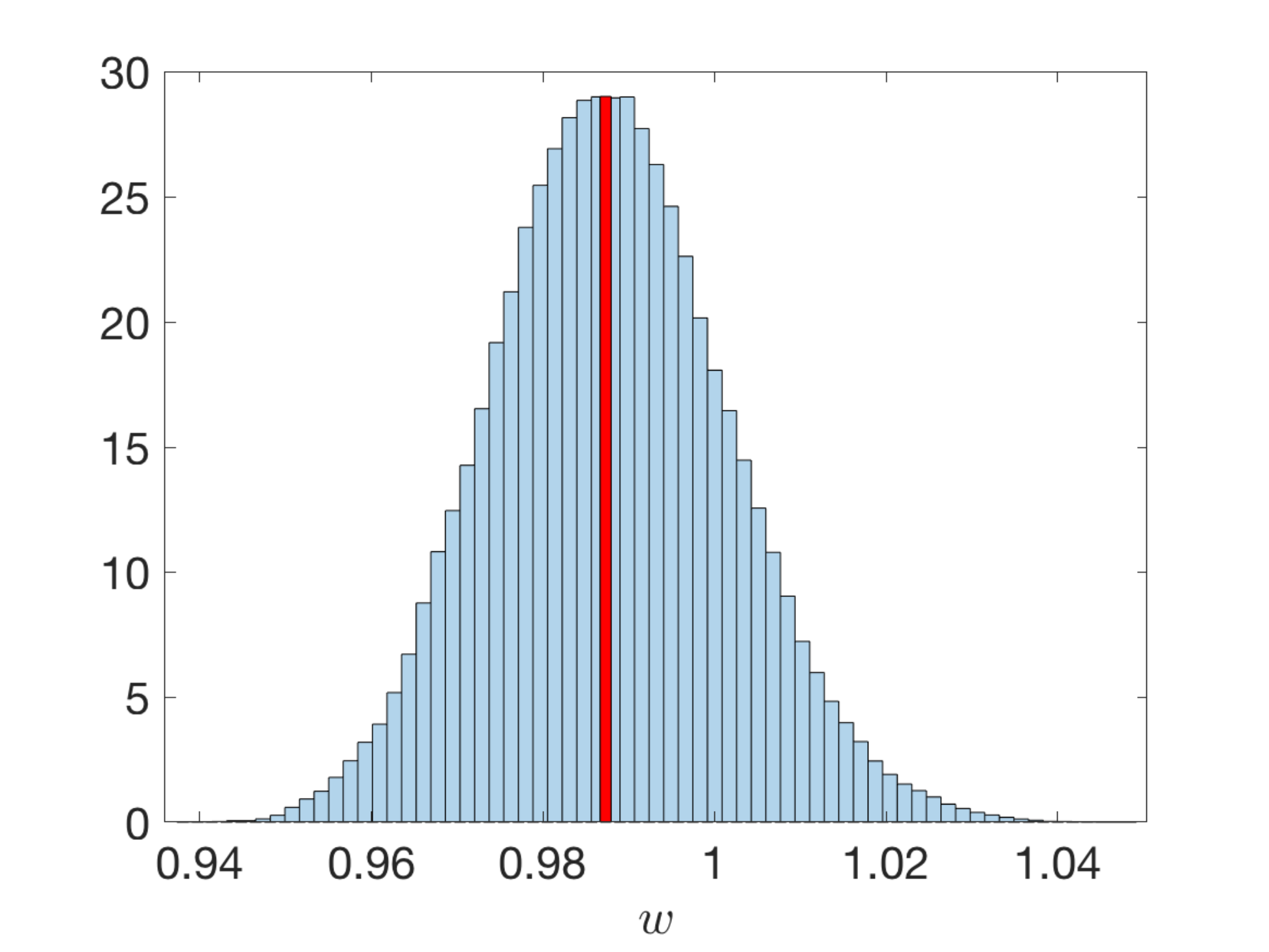}
\includegraphics[width = 0.35\columnwidth]{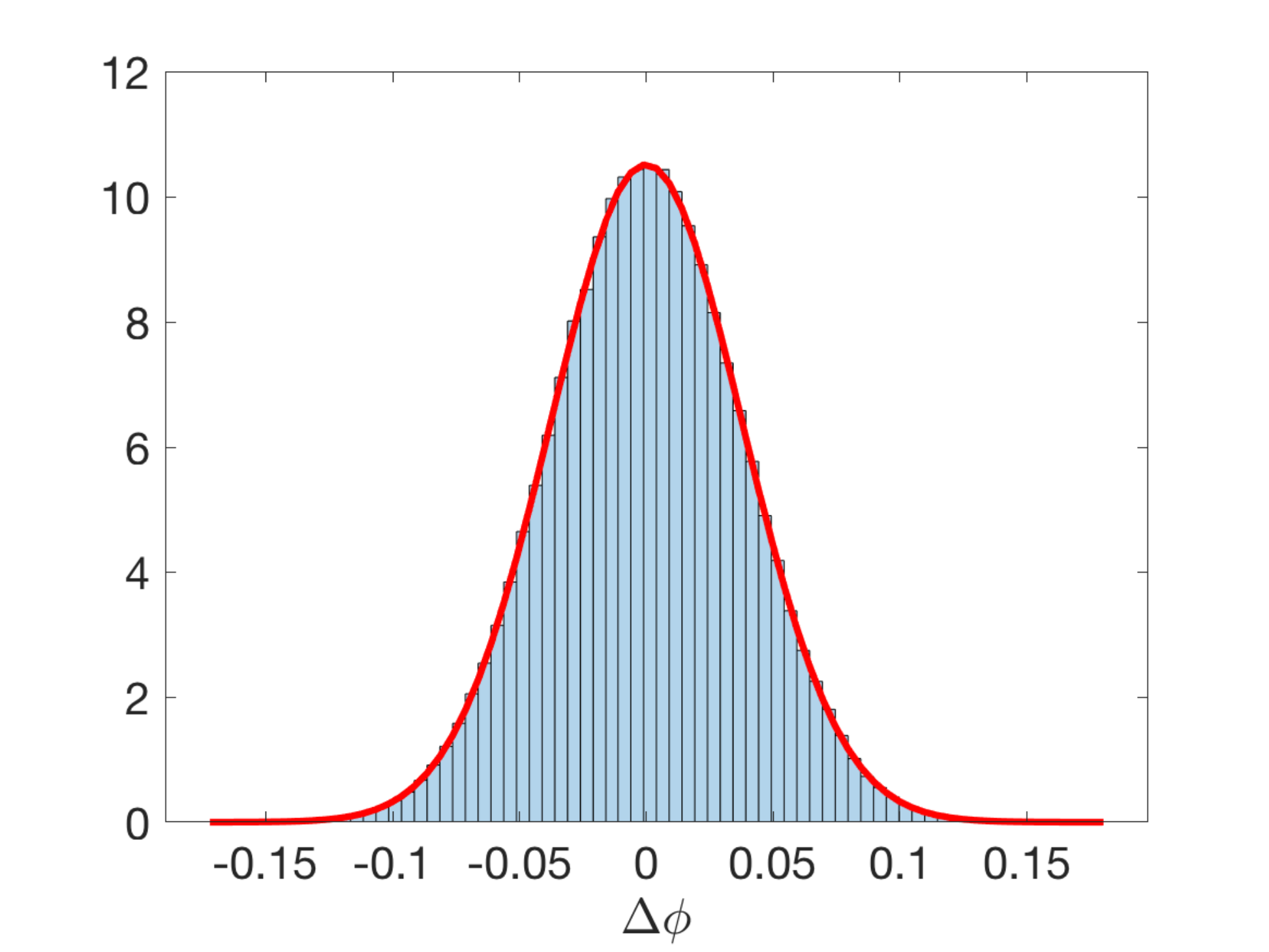}
\caption{Left: Empirical histogram of the inverse front width $w$ for the bistable SPDE (\ref{e.bistable}) with multiplicative noise. The deterministic equilibrium inverse width of the collective coordinates approach (\ref{e.bistable_mult_w}) is indicated in red. Right: Empirical histogram of the difference $\Delta \phi$ between the interface location $\phi$ at successive times $=n\Delta t$. The implied analytical probability density function according to the collective coordinate approach (\ref{e.SDE_bistable_mult}) is depicted in red. The histograms were obtained from a long simulation of $5000$ time units with parameters $D=0.2$ and $b=0.1$, $\sigma=0.75$.}
\label{f.bistable_mult}
\end{figure}

We now consider the case when the stochastic forcing of the SPDE is different from the stochastic forcing of the equations for the collective coordinates and $W(t) \neq B(t)$. We estimate the statistical behaviour of the inverse width of the travelling wave and its speed by computing temporal averages. We have checked that the chosen integration time $T=5000$ time units is sufficiently long to allow for converged statistics. To ensure that the travelling wave remains within the domain during the time of integration we move into the frame of reference co-moving with speed $\bar c_0=c_0/\sqrt{1+\sigma^2}$.\\ Figure~\ref{f.bistable_mult} shows the empirical histogram of the inverse width of the travelling wave front of the bistable SPDE (\ref{e.bistable}) with multiplicative noise. The mean $\E[w]=0.9878$ is approximated by the collective coordinate solution $\bar w = \sqrt{1+\sigma^2}\, w_0=0.9882$ (cf. (\ref{e.bistable_mult_w})), remarkably well with a relative error of $0.046\%$. The inverse width $w$ experiences small fluctuations with $\V[w]=1.9\, 10^{-4}$. We have checked that the variance decreases with increasing length of the time series $T$, consistent with the analytical result $\V[w]=0$ obtained from the collective coordinates.  

The average speed of the front is estimated as $\lim_{t\to\infty} \phi/t\approx 0.2036$ and matches the result of the collective coordinates with its prediction $\lim_{t\to\infty} \phi/t = c_0/\sqrt{1+\sigma^2} = 0.2024$ up to a relative error of $0.59\%$. The fluctuations around the linear drift can be estimated via the variance of $\Delta \phi$ (after subtracting the mean linear drift). We find $\V[\Delta \phi]/\tau=0.143$ is well approximated by the collective coordinate prediction $\V[\Delta\phi]/\tau = \sigma^2/(2\bar w)^2=0.144$ with a relative error of $0.04\%$. We have also estimated the mean-square displacement $\text{MSD}(t)=\lim_{T\to\infty}\frac{1}{T}\int_0^T\phi(t+s)\phi(s)\,ds=\sigma_{\phi\phi}^2t+o(1)$, confirming the accuracy of the collective coordinate approach (not shown). Figure~\ref{f.bistable_mult} shows the empirical histogram of $\Delta\phi = \phi(t+\tau)-\phi(\tau)$ for $\tau=dt$. Recall that we solve the SPDE (\ref{e.bistable}) in the frame of reference moving with speed $\bar c_0=c_0/\sqrt{1+\sigma^2}$. The histogram is then centred at zero by subtracting the residual linear drift of the location of the front using linear regression. The figure clearly shows that the empirical histogram is very well approximated by the prediction of the collective coordinates with a Gaussian with mean zero and variance $\sigma^2/(4 \bar w^2)$.

For $b=0.5$ the deterministic speed is $c_0=0$; the collective coordinate ansatz predicts that the noise does not induce any non-zero speed corrections and has $\bar c_0=0$ (cf (\ref{e.bistable_mult_c0})). This is confirmed in simulations of the full SPDE with an average speed of $0.005$; the estimated value of the average speed decreases with the length $T$ of the simulation time.

\section{Collective coordinate approach for the bistable Nagumo equation with additive noise}
\label{s.bistable_add}
Consider the bistable SPDE with additive noise defined in (\ref{e.QWiener})
\begin{align}
\partial_t u = D \partial_{xx}u + u(1-u)(u-b) + \sigma s(x)\sum_{j=1}^\infty\sqrt{\lambda_j}\varphi_j(x) {\dot{B}}_j ,
\label{e.SPDEbistable_add}
\end{align}
driven by a $Q$-Wiener process with kernel of the covariance operator ${\mathcal{C}}(x,y)=e^{-|x-y|/\ell}$ supported on a finite domain with $s(x)=1$ for $x\in [-\tfrac{1}{2}L_{\rm{noise}},\tfrac{1}{2}L_{\rm{noise}}]$ and $s(x)=0$ otherwise (which we approximate by the smooth function for $s(x)$ given by (\ref{e.noiselocal}) using $\kappa\gg 1$). $\lambda_j$ and $\varphi_j$ represent the eigenvalues and eigenfunctions of the covariance operator, satisfying
\begin{align}
\int {\mathcal{C}}(x,y)\varphi_j(y) \, dy = \lambda_j \varphi_j(x).
\end{align}
For ${\mathcal{C}}(x,y)=e^{-|x-y|/\ell}$ the eigenvalue problem can be solved analytically on a finite domain of length $L$ \cite{Ghanem}, and we have 
\begin{align}
\varphi_j(x)
 = 
\begin{cases}
\sqrt{
\frac{2p_j}{
p_j L
-
\sin\left(p_j L\right)}}
\sin\left(p_j x\right) 
& {\rm{if}}\;  j \;{\rm{is \;even}}\\
\sqrt{
\frac{2q_j}{
q_j L
+
\sin\left(q_j L\right)}}
\cos\left(q_j x\right) 
&   {\rm{if}}\; j\; {\rm{is \;odd}}
\end{cases}
\label{e.varphi}
\end{align}
with 
eigenvalues
\begin{align}
\lambda_j=
\begin{cases}
\frac{2\ell}{1+\ell^2 p_j^2} & {\rm{if}}\;  $j$ \;{\rm{is \;even}}\\
\frac{2\ell}{1+\ell^2 q_j^2}&   {\rm{if}}\; $j$\; {\rm{is \;odd}}
\end{cases},
\label{e.lambda}
\end{align}
where $p_j$ and $q_j$ are the solutions of the transcendental equations
\begin{align*}
\ell\, p + \tan(p\tfrac{L}{2}) &=0 \qquad {\rm{for \; even }}\;  j\\
1 -\ell \, q \tan(q\tfrac{L}{2}) &=0 \qquad {\rm{for \; odd }}\;  j.
\end{align*}

The addition of spatially localized noise breaks the translational invariance. We employ again the $\tanh$-ansatz function (\ref{e.cc}) as for the case of multiplicative noise and assume that the dynamics of the collective coordinates is given by the system of SDEs 
\begin{align}
d w & = a_w(w,\phi)\, dt + \sum_{k=0}^\infty \sigma_{w k}(w,\phi)\, dW_k(t)
\label{e.w_add}
\\
d\phi &=  a_\phi(w,\phi)\, dt + \sum_{k=0}^\infty \sigma_{\phi k}(w,\phi)\, dW_k(t).
\label{e.phi_add}
\end{align}
This is only justified for small noise amplitude $\sigma$ and sufficiently small noise regions $L_{\rm{noise}}$, where the probability of noise induced nucleation of new fronts is small. We  obtain again the projections (\ref{e.proj_w_1}) and (\ref{e.proj_phi_1}), determining the drift coefficients $a_w$ and $a_\phi$, which we can evaluate (using the explicit integrals given in \ref{app.1}) as
\begin{align}
a_w &= -\frac{3}{4}\frac{w}{w_0^2(\pi^2-6)}\left(w^2-w_0^2 \right)
+ \frac{3}{4w}\sum_{k=1}^\infty\sigma_{w k}^2  + \frac{3w^3}{\pi^2-6}\sum_{k=1}^\infty\sigma_{\phi k}^2,
\label{e.aw_additive}
\\
a_\phi & = \frac{1}{4w}(1-2b) - \frac{1}{2w}
\sum_{k=1}^\infty\sigma_{wk}\sigma_{\phi k} .
\label{e.aphi_additive}
\end{align}
The projections onto the stochastic terms yield, upon using $\langle \pdif{u}{w}\pdif{u}{\phi}\rangle =0$,
\begin{flalign}
&
\langle \left(\pdif{u}{w}\right)^2\rangle\, 
\sum_{k=1}^\infty\sigma_{w k}\, dW_k(t)
=
\sigma \sum_{j=1}^\infty \sqrt{\lambda_j} \langle s \pdif{u}{w} \varphi_j \rangle\, dB_j(t)
\label{e.diff_proj_w_additive}
\\
&\langle \left(\pdif{u}{\phi}\right)^2\rangle \,
\sum_{k=1}^\infty\sigma_{\phi k}\, dW_k(t)
=
\sigma \sum_{j=1}^\infty \sqrt{\lambda_j}\langle s \pdif{u}{\phi} \varphi_j \rangle\, dB_j(t) .
\label{e.diff_proj_phi_additive}
\end{flalign}
Again pathwise matching can be achieved for $W_k=B_k$ with diffusion coefficients
\begin{align}
\sigma_{w k} = \sigma \sqrt{\lambda_k}\frac{\langle s \frac{\partial u}{\partial w}\varphi_k\rangle}{\langle \left(\frac{\partial u}{\partial w}\right)^2\rangle}\qquad {\rm{and}} \qquad 
\sigma_{\phi k} = \sigma \sqrt{\lambda_k}\frac{\langle s \frac{\partial u}{\partial \phi}\varphi_k\rangle}{\langle \left(\frac{\partial u}{\partial \phi}\right)^2\rangle}\, .
\label{e.sigma_add_path}
\end{align}
Note that when the front interface is well outside of the domain $[-\tfrac{1}{2}L_{\rm{noise}},\tfrac{1}{2} L_{\rm{noise}}]$, where $s\equiv 0$, the collective coordinates satisfy essentially deterministic evolution equations with $\sigma_{w k} =\sigma_{\phi k} =0$ for all $k$. The breaking of the translational symmetry by the localized additive noise now allows for non-deterministic diffusive behaviour of the shape variable $w$.\\ 

To avoid having to deal with infinitely many diffusion coefficients, we now present the calculations when truncating to $k\le 2$ in (\ref{e.w_add})--(\ref{e.phi_add}), matching the mean and the variance of the now two-dimensional Brownian motion $W_{1,2}$ of the collective coordinates with the statistics of the $\infty$-dimensional $Q$-Wiener process. The truncation to $k\le 2$ is justified for Gaussian or near-Gaussian stochastic forcing, which we will see is indeed the case for our example of the bistable equation (\ref{e.SPDEbistable_add}) driven by additive noise. We introduce
\begin{align*}
S
= \left(
\begin{matrix}
 \langle \left(\pdif{u}{w}\right)^2\rangle \sigma_{w1} &  \langle \left(\pdif{u}{w}\right)^2\rangle \sigma_{w 2}\\
 \langle \left(\pdif{u}{\phi}\right)^2\rangle \sigma_{\phi 1} &  \langle \left(\pdif{u}{\phi}\right)^2\rangle \sigma_{\phi 2}
 \end{matrix}
 \right)
\end{align*}
and
\begin{align*}
G
= \left(
\begin{matrix}
 \sqrt{\lambda_1} \langle s \pdif{u}{w} \varphi_1\rangle &  \sqrt{\lambda_2} \langle s \pdif{u}{w} \varphi_2\rangle & \cdots & \\
 \sqrt{\lambda_1} \langle s \pdif{u}{\phi}\varphi_1\rangle &  \sqrt{\lambda_2} \langle s \pdif{u}{\phi} \varphi_2\rangle & \cdots & \\
 \end{matrix}
 \right),
\end{align*}
and, as outlined in Section~\ref{s.cc}, require that the implied diffusion coefficients of the two stochastic processes in 
(\ref{e.diff_proj_w_additive})--(\ref{e.diff_proj_phi_additive}) are equal with 
\begin{align}
SS^T = GG^T .
\label{e.SS}
\end{align}
The diffusion coefficients of the collective coordinates can then be determined as a square root, and give rise to multiplicative noise. 
%
%
%
%
%
%
%
From (\ref{e.SS}) we can immediately derive
\begin{align}
\sigma^2_{w1}+\sigma^2_{w2} & =  \frac{\sigma^2}{\langle \left(\pdif{u}{w}\right)^2\rangle^2} 
\sum_{j=1}^\infty  \lambda_j \langle s \pdif{u}{w} \varphi_j \rangle^2 
\label{e.add_diff_w} 
\\
\sigma^2_{\phi 1}+\sigma^2_{\phi 2} & =  \frac{\sigma^2}{\langle \left(\pdif{u}{\phi}\right)^2\rangle^2} 
\sum_{j=1}^\infty  \lambda_j \langle s \pdif{u}{\phi} \varphi_j \rangle^2 
\label{e.add_diff_phi}
\\
\sigma_{w1}\sigma_{\phi 1}+\sigma_{\phi 2}\sigma_{w 2} & =  \frac{\sigma^2}{\langle\left( \pdif{u}{w}\right)^2\rangle\langle \left(\pdif{u}{\phi}\right)^2\rangle} 
\sum_{j=1}^\infty  \lambda_j \langle s \pdif{u}{w} \varphi_j  \rangle \langle s \pdif{u}{\phi} \varphi_j \rangle ,
\label{e.add_diff_mix}
\end{align}
which affect the drift coefficients  (\ref{e.aw_additive})--(\ref{e.aphi_additive}) and also determine the variance of increments of the collective coordinates with $\V[dw]/dt= \sigma^2_{w1}+\sigma^2_{w2} $ and $\V[d\phi]/dt=\sigma^2_{\phi 1}+\sigma^2_{\phi 2}$ (upon subtracting the linear drift from $\phi$).\\ 

Since the diffusion coefficients depend on $w$ and $\phi$ we are not able to find an explicit solution for the expected value of the inverse width $\E[w]$ by solving for $a_w\equiv 0$ in (\ref{e.aw_additive}). However, for $\sigma \ll 1$ we can approximate $\E[w]\approx w_0$. Similarly we find that $\E[\phi] \approx \phi_0+c_0 t$ (cf (\ref{e.aphi_additive})) in the small noise limit. Recall that we required $\sigma\ll 1$ in order to prevent nucleation of new fronts within the noise region during the time of the front passing through. 


\subsection{Numerical results}
We now present the numerical results for the case of additive noise and compare results from simulating the Nagumo equation (\ref{e.SPDEbistable_add}) with additive noise with the analytical results from the collective coordinate approach where we employ here the expressions for the diffusion coefficients (\ref{e.sigma_add_path}) yielding pathwise approximation. We present results for $D=0.1$ and $b=0.25$. To define the $Q$-Wiener process we choose $\sigma=0.022$ and a spatial correlation length of $\ell=0.25$. The noise is located in a region centred around $x=0$ of spatial extension $L_{\rm{noise}}=5$ and we choose $\kappa=5$ for the localization function $s(x)$ (cf. (\ref{e.noiselocal})). The $Q$-Wiener process is constructed employing a circulant embedding of the covariance following the construction in \cite{Lord}. This method produces complex valued noise, of which the real and imaginary part are independent and identically distributed (only the real part is used in our simulations). The infinite-sum of the $Q$-Wiener process in (\ref{e.SPDEbistable_add}) is truncated to $M=191$ where $\lambda_{191}=0.069$. An outline of the method is included in \ref{app.2}. The SPDE (\ref{e.SPDEbistable_add}) with additive noise is then solved using an Euler-Maruyama scheme whereby the diffusion operator is treated implicitly. We choose as the domain $x\in [-30,30]$ of length $L=60$ and employ a temporal discretization step of $\Delta t=0.0025$ and a spatial discretization with grid spacing $\Delta x = 0.05$. The width of the travelling front and the location of the interface $\phi(t)$ where the fitted $\tanh$-profile assumes the value $\hat u\equiv 0.5$ are determined using the same method as for multiplicative noise. We remark that, unlike in the case of multiplicative noise, for additive noise the front  solution of the SPDE is stochastically perturbed around $u=0$ and $u=1$ and the location of the front interface with $u(\xi)=0.5$ is not necessarily unique. 
We perform a total of $500$ independent realizations for the SPDE (\ref{e.SPDEbistable_add}). Each realization of the SPDE simulates a full traverse of the front across the noise region $[-\tfrac{1}{2}L_{\rm{noise}},\tfrac{1}{2}L_{\rm{noise}}]$. We checked that no additional fronts have nucleated within this time. We further perform a total of $500$ realizations of the collective coordinate SDEs (\ref{e.w_add})--(\ref{e.phi_add}) with drift coefficients (\ref{e.aw_additive})--(\ref{e.aphi_additive}) and diffusion coefficients estimated from (\ref{e.sigma_add_path}) with $k\le 191$. We have also performed simulations of the system of SDEs for the collective coordinates (\ref{e.w_add})--(\ref{e.phi_add}) with diffusion coefficients estimated from (\ref{e.SS}) (i.e. $k\le 2$), assuming Gaussian stochastic forcing (not shown). The two expressions of the diffusion matrix using either $k\le 191$ to obtain pathwise correspondence or $k\le 2$ yield very similar results.

\begin{figure}[htbp]
\centering
\includegraphics[width = 0.35\columnwidth]{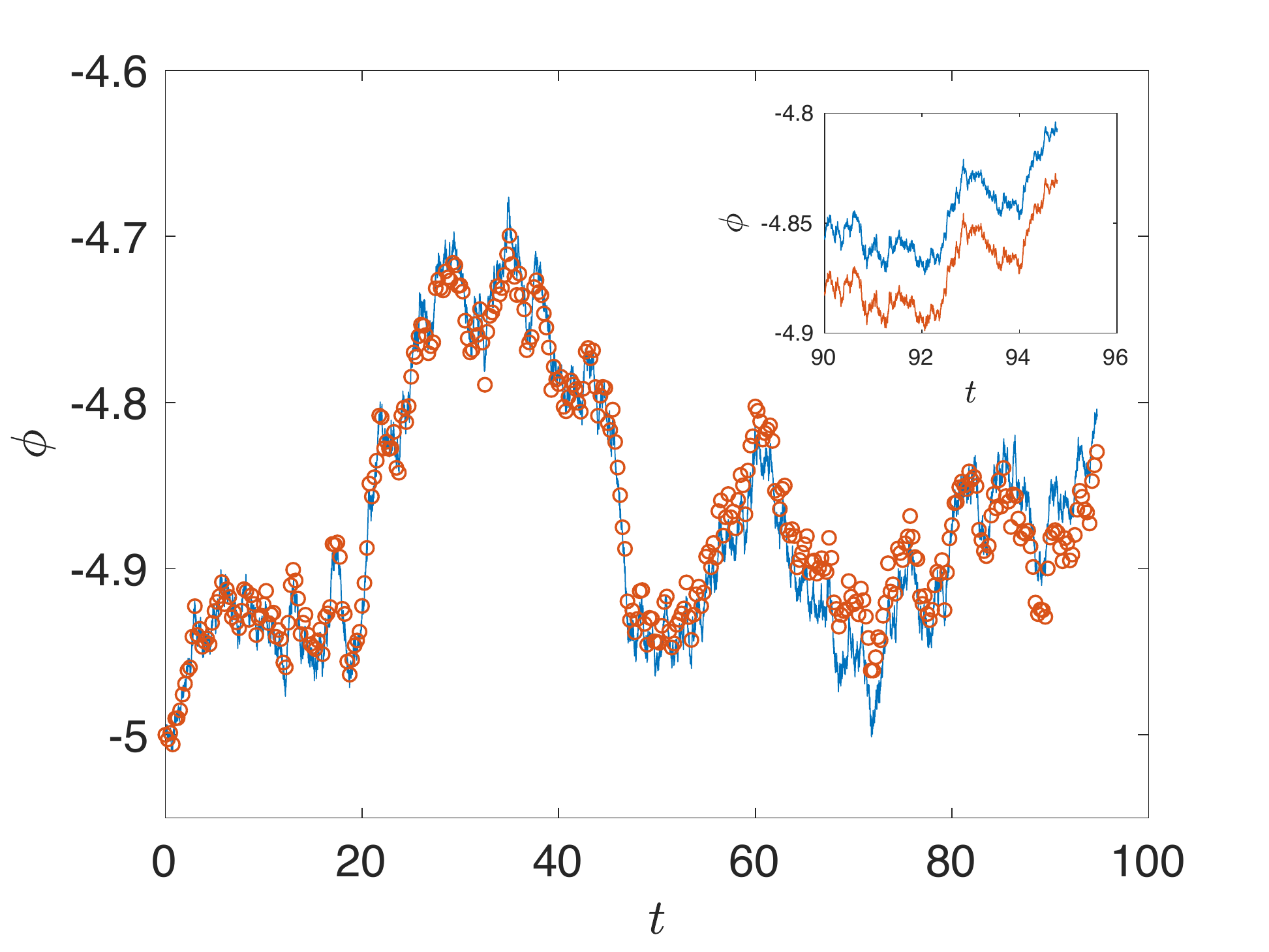}
\includegraphics[width = 0.35\columnwidth]{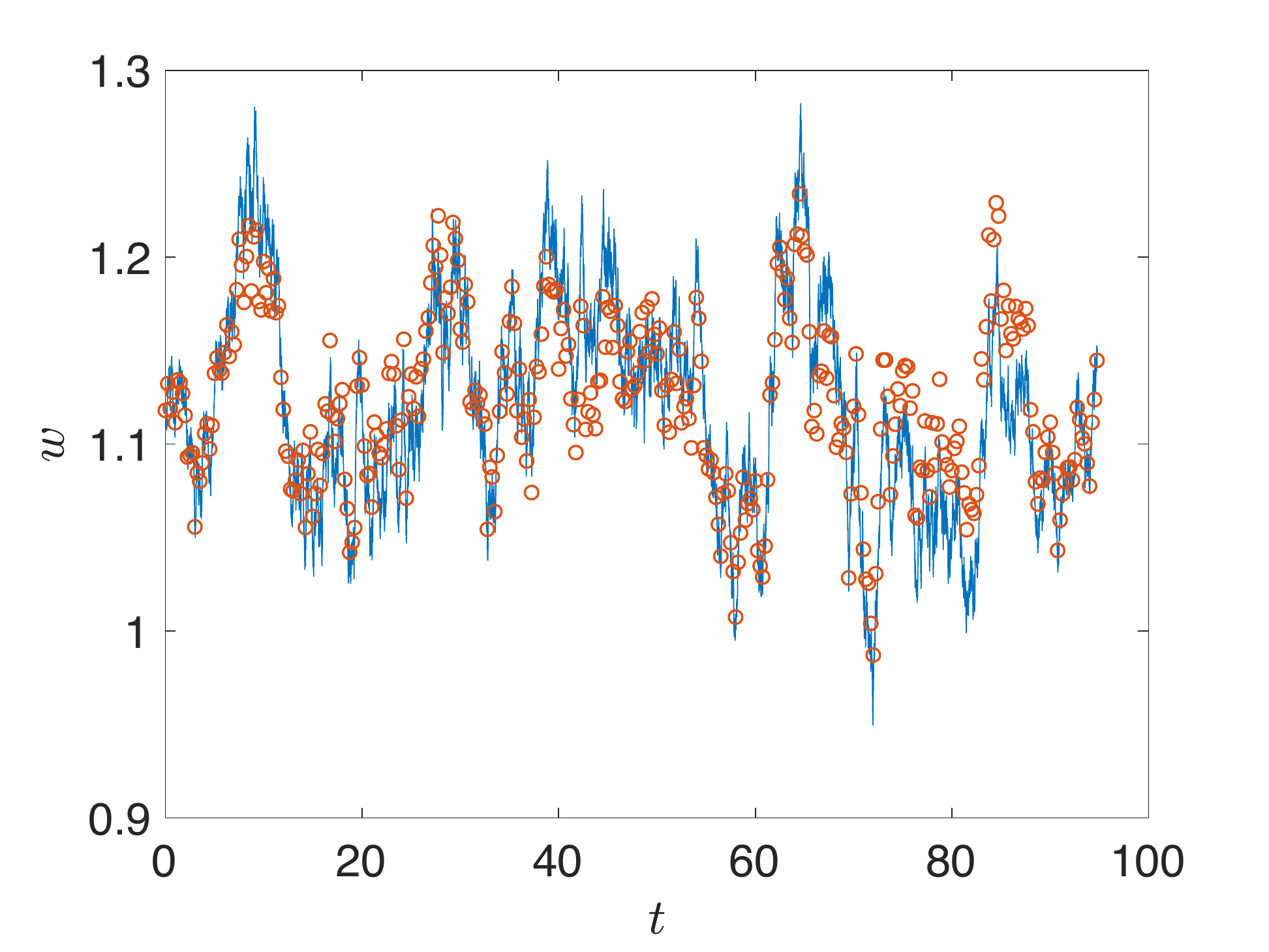}
\caption{Comparison of the time series of the location of the interface $\phi(t)$ (left) and the inverse front width $w$ (right) for a travelling front solution for the bistable equation (\ref{e.SPDEbistable_add}) with $D=0.1$ and $b=0.25$ driven by additive noise with $\sigma=0.022$, $L_{\rm{noise}}=5$, $\ell=0.25$ and $\kappa=5$. Shown are solutions obtained from a simulation of the bistable SPDE for one realization of the stochastic forcing $B(t)$ (blue continuous line) as well as the solution of the collective coordinate equations in the pathwise case  $W(t)=B(t)$ using (\ref{e.aw_additive})--(\ref{e.aphi_additive}) and (\ref{e.sigma_add_path}) (red circles; evaluated at every $100$ time steps to allow for better comparison).} 
\label{f.bistable_add_pathwise}
\end{figure}

Figure~\ref{f.bistable_add_pathwise} shows the phase $\phi(t)$ and the inverse width $w(t)$ of a travelling front obtained from a simulation of the bistable SPDE (\ref{e.SPDEbistable_add}) with additive noise as well the respective collective coordinate approximations provided by the system of SDEs (\ref{e.w_add})--(\ref{e.phi_add}) with the drift coefficients given by (\ref{e.aw_additive})--(\ref{e.aphi_additive}) and diffusion coefficients given by (\ref{e.sigma_add_path}) when the same stochastic forcing is used for both systems with $W(t)=B(t)$. Results are shown during a full traverse of the noisy region; the entrance of the front is marked by setting $t=0$ and the exit occurs at $t\approx 95$.  
It is seen that the pathwise approximation is very good when the front is entering the noisy region close to $t=0$ as well as when it is exiting the noisy region close to $t=95$; at exit $w(t)$ is very well approximated and the difference in the phase is a constant shift induced by the previously accumulated error (see the inset in Figure~\ref{f.bistable_add_pathwise}). The maximum relative errors for $w$ and in $\phi$ over the whole traverse are $4\%$ and $0.7\%$, respectively, and the average relative error is only $1.6\%$ for $w$ and $0.2\%$ for $\phi$. Near $t=0$ and $t=95$ where the unperturbed trailing and leading tail of the front is conducive to fitting a $\tanh$-profile to the stochastically modified front solution, the approximation is much better, suggestive of the effectiveness of our approximation to provide pathwise approximation. The error inside the domain can be explained by the deficiency of the employed nonlinear least square fitting routine for $\tanh$ profiles which are contaminated by spatially correlated noise with a finite correlation length $\ell$. This is consistent with the remarkably good agreement near the entrance and near the exit of the front.
%
\\

\begin{figure}[htbp]
	\centering
	\includegraphics[width = 0.35\columnwidth]{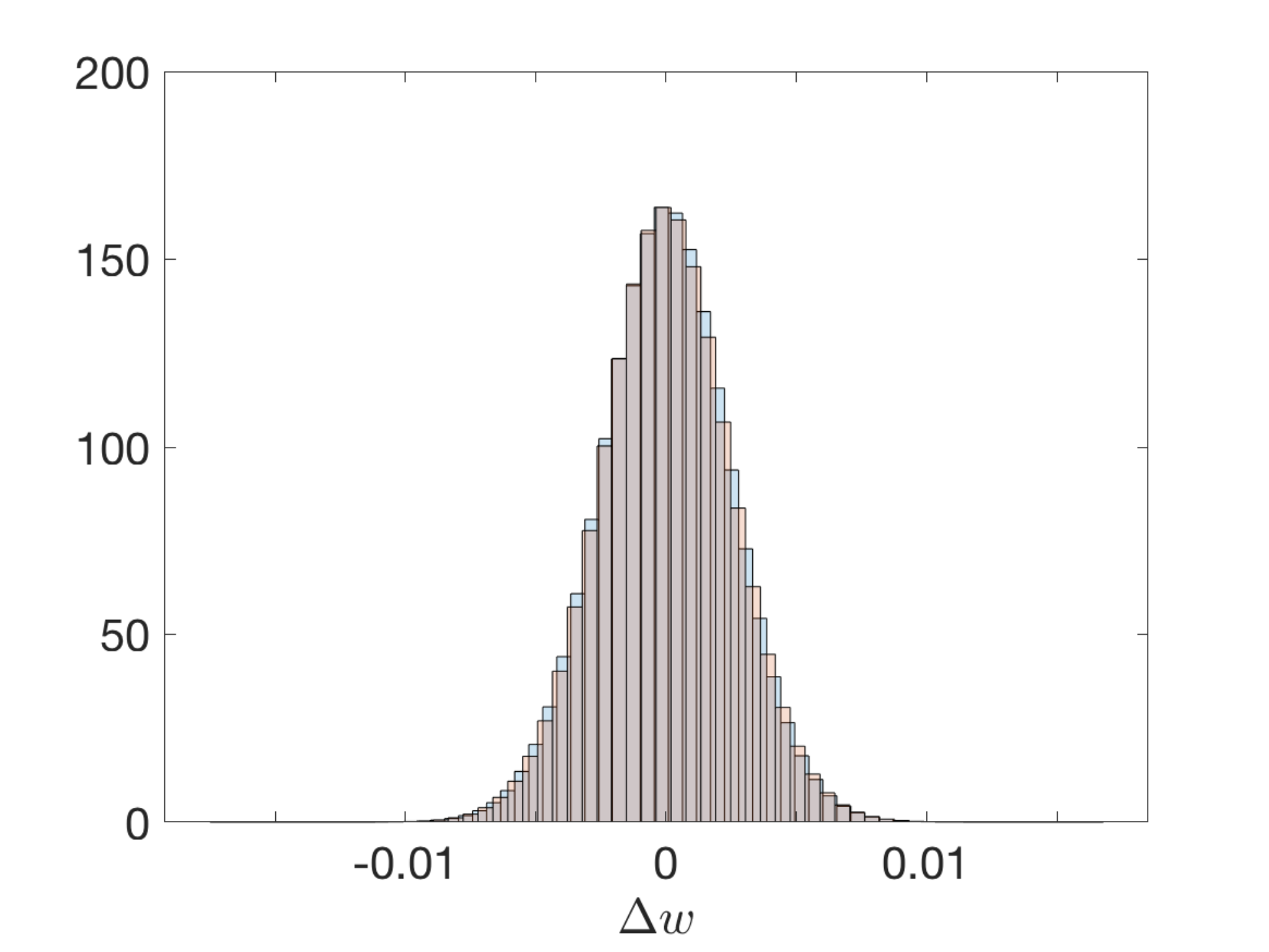}
	\includegraphics[width = 0.35\columnwidth]{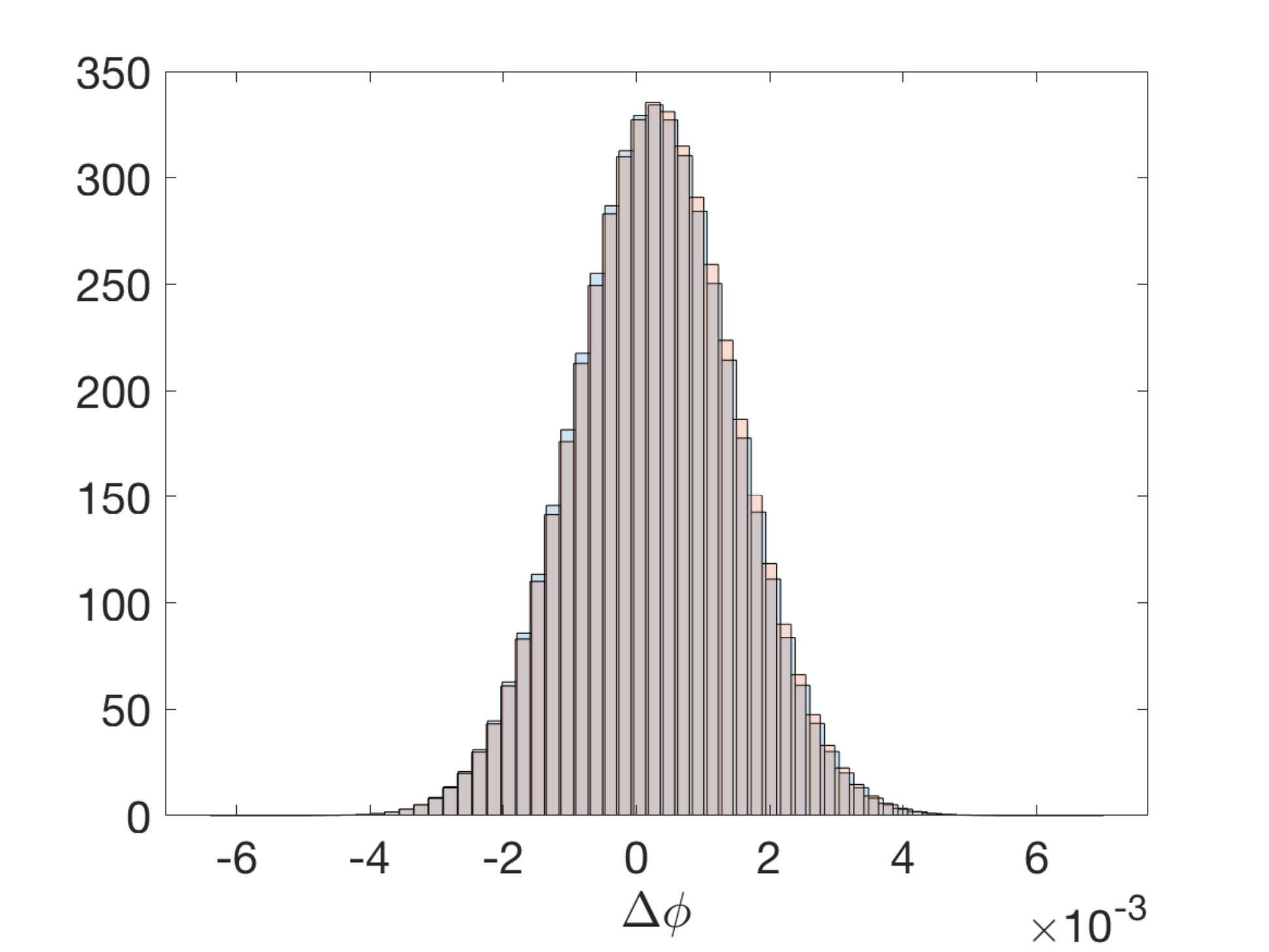}
	\caption{Marginal distributions for the bistable SPDE (\ref{e.SPDEbistable_add}) with additive noise and their predictions from the collective coordinate approach. The distributions for the full SPDE are given in blue and the distributions of the collective coordinates approach are superimposed in red. Left: Empirical histograms for the inverse width increments $\Delta w$. Right: Empirical histogram of the difference $\Delta \phi$ between the interface location $\phi$ at successive times. The mean drift $c_0\,t$ is subtracted from $\phi(t)$ to centre the  histogram at $0$. The histograms were obtained from $500$ realizations, recording $w$ and $\phi$ when the front is passing through the noisy region. Parameters are as in Figure~\ref{f.bistable_add_pathwise}.}
	\label{f.bistable_add}
\end{figure}

\begin{figure}[htbp]
	\centering
	\includegraphics[width = 0.35\columnwidth]{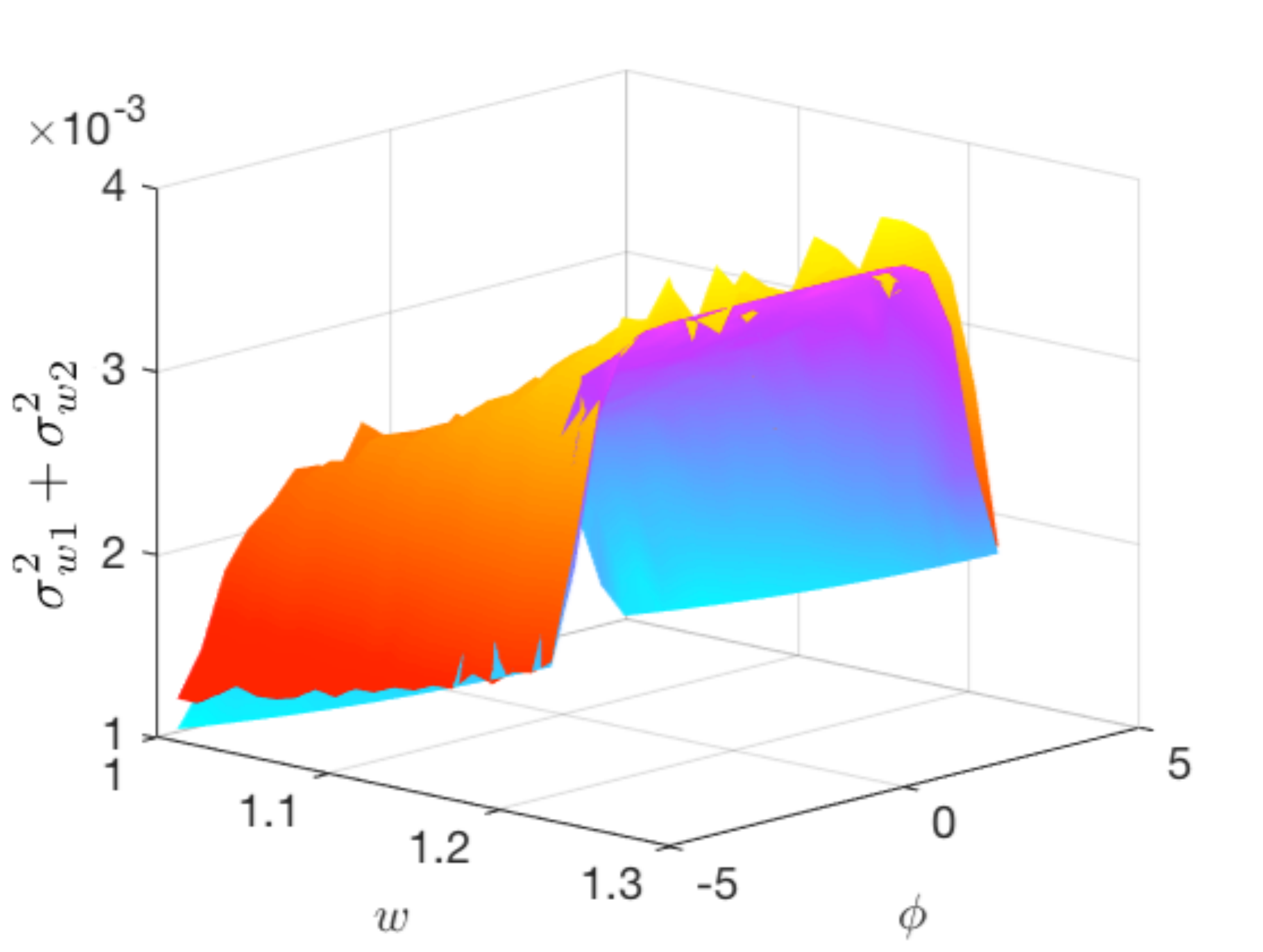}
	\includegraphics[width = 0.35\columnwidth]{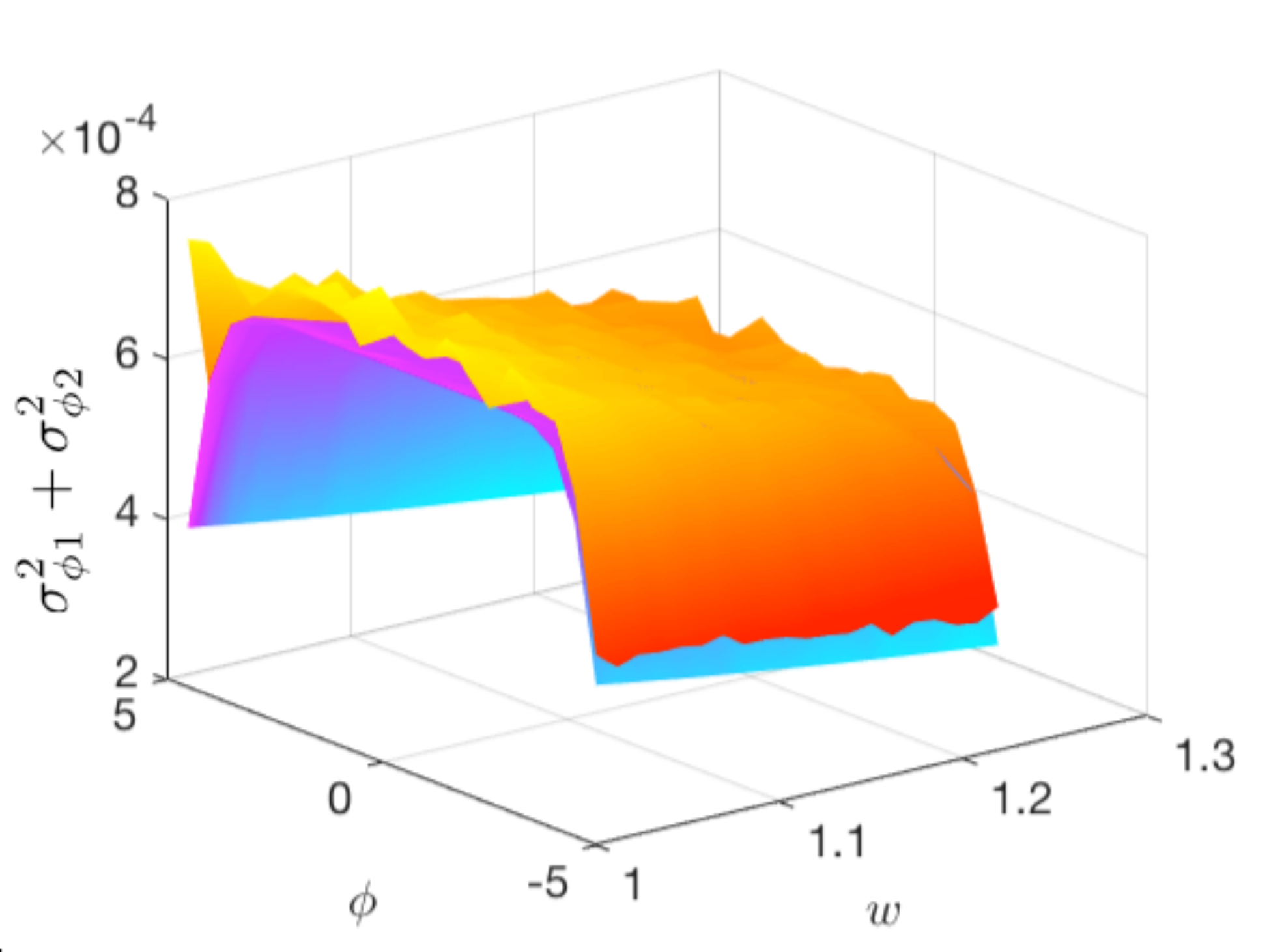}\\
	\includegraphics[width = 0.35\columnwidth]{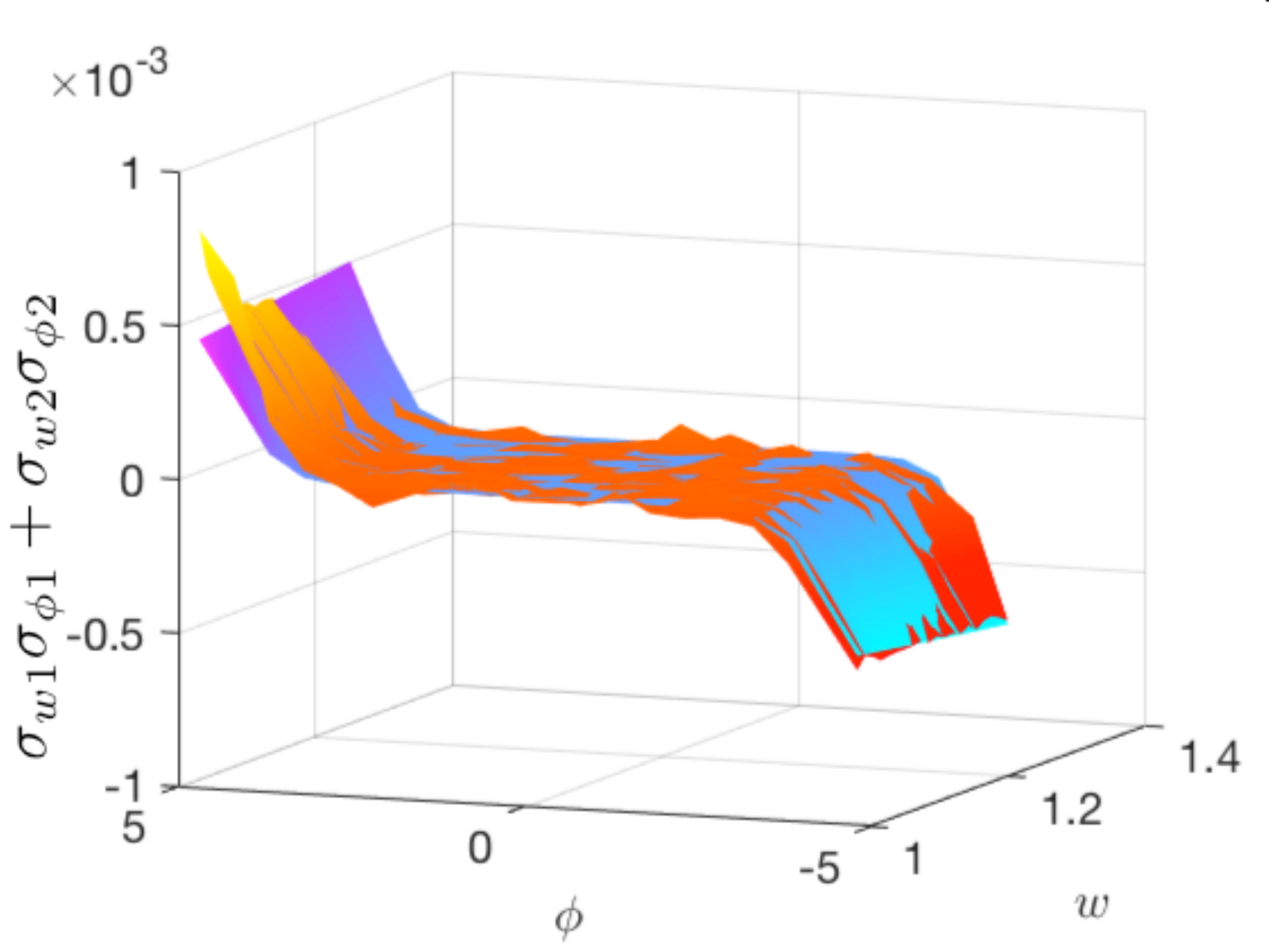}
	\caption{Variances $\V[\Delta w]/{\Delta t}$ and $\V[\Delta \phi]/\Delta t$ as well as $\E[\Delta w\Delta\phi]/\Delta t$, as functions of $w$ and $\phi$. The results estimated from the bistable SPDE (\ref{e.SPDEbistable_add}) with additive noise are shown in orange. The corresponding analytical results $\sigma_{w1}^2+\sigma_{w2}^2$, $\sigma_{\phi 1}^2+\sigma_{\phi 2}^2$ and $\sigma_{w1}\sigma_{\phi 1}+\sigma_{w2}\sigma_{\phi 2}$ respectively calculated from (\ref{e.add_diff_w})--(\ref{e.add_diff_mix}) using the collective coordinate approach are shown in purple/blue. Parameters as in Figure~\ref{f.bistable_add_pathwise}.}
	\label{f.bistable_add_diff}
\end{figure}

We consider now the case when the stochastic forcing of the SPDE is different from the stochastic forcing of the equations for the collective coordinates and $W(t) \neq B(t)$. The mean inverse width of the full SPDE is estimated as $\mathbb{E}[w]=1.116$, and is well approximated by the deterministic limit of the collective coordinates approach $\mathbb{E}[w]\approx  w_0=1.118$ averaged over $500$ realizations. The approximation is accurate with a relative error of only $0.18\%$. Similarly, the average front propagation speed is estimated from the SPDE with $\E[\phi] = \phi_0+0.1123\, t$ whereas the collective coordinate formulae yield $\E[\phi] = \phi_0+ 0.1116\, t$ with a relative error of $0.7\%$. We found that due to the localized nature of the noise, the nonlinear least square fitting is not adequate to determine the inverse front width $w$, when fitting solutions of the form $\hat u(x;w,\phi)=\ha (1-\tanh(w(x-\phi)))$. We have checked this against artificially generated fronts with randomly chosen $w$ and $\phi$ and found that $\V[w]$ is approximately ten times larger than the specified variance of the artificially generated front ensemble when using a nonlinear least square fit. The mean $\E[w]$, however, was correctly estimated. 
We therefore present in Figure~\ref{f.bistable_add} and in Figure~\ref{f.bistable_add_diff} the empirical histograms for the increments $\Delta w$ of the inverse width, as well as for $\Delta\phi$, conditioned on the position of the front being within the region of noise with $-\tfrac{1}{2}L_{\rm{noise}}\le \phi\le \tfrac{1}{2}L_{\rm{noise}}$. The marginal distributions are depicted in Figure~\ref{f.bistable_add} and the fluctuations around the mean inverse width $w$ and around the linear drift of the front location as functions of $w$ and $\phi$ are depicted in Figure~\ref{f.bistable_add_diff}. The marginal distributions of the increments of the inverse width and the front location are very well approximated by the collective  coordinate approach (see Figure~\ref{f.bistable_add}). The variance of the increments of the inverse width are estimated with 
$\V[\Delta w ]=2.482\cdot10^{-3}\, \Delta t$ for the SPDE and with  
$\V[\Delta w]=2.499\cdot10^{-3}\, \Delta t$ for the collective coordinates, with a relative error of $0.7\%$. Similarly, the variance $\V[\Delta \phi]=5.834\cdot10^{-4}\,\Delta t$ for the SPDE is reproduced well by the collective coordinate approach which predicts $\V[\Delta \phi]= 5.793\cdot10^{-4}\, \Delta t$, implying a relative error of $0.7\%$. 
The fluctuations around the mean inverse width $w$ and around the linear drift of the front location $\phi$, $\V[dw/\sqrt{dt}]= \sigma^2_{w1}+\sigma^2_{w2} $, $\V[d\phi/\sqrt{dt}]=\sigma^2_{\phi 1}+\sigma^2_{\phi 2}$ as well as $\sigma_{w1}\sigma_{\phi 1}+\sigma_{w 2}\sigma_{\phi 2}$, are shown in Figure~\ref{f.bistable_add_diff} as functions of $w$ and $\phi$. It is seen that the predictions (\ref{e.add_diff_w})--(\ref{e.add_diff_mix}) of the collective coordinate approach capture the nontrivial dependency on $w$ and $\phi$ of the diffusive behaviour of the travelling front very well. 

\section{Discussion and outlook}
\label{s.discussion}
Employing a symmetry perspective whereby the dynamics of a travelling front is split into the shape dynamics and the dynamics along the symmetry group, we introduced a collective coordinate framework to reduce an infinite dimensional SPDE to a system of finite dimensional SDEs. This allowed us to quantitatively describe the behaviour of travelling fronts in the presence of noise. A crucial assumption was that the shape dynamics is strongly contracting, which prevents the noise from strongly affecting the shape of the travelling front with respect to its deterministic solution. We studied the particular example of a stochastic bistable Nagumo equation. In the case of multiplicative noise, the effective reduced SDE describes a deterministic front shape with a constant width and diffusive behaviour along the translational group of the spatial location of the front.  We found that multiplicative noise slows down the travelling front and leads to a sharper front interface. We further investigated the ability of the collective coordinate framework to capture the diffusive behaviour of travelling waves when the translational symmetry is broken, as in the case of additive spatially localized noise. In both cases the collective coordinate approach allowed a quantitative description of the statistical features of the dynamics of the travelling front remarkably well. Whereas in the case of multiplicative noise only the dynamics along the neutral group is found to be diffusive, the collective coordinate approach finds diffusive behaviour of both the group and the shape dynamics in the case of the symmetry breaking additive noise.\\

It is pertinent to mention that the actual form of the ansatz function (\ref{e.cc}) is not relevant. We have also performed the calculation for an ansatz function $\hat u(x,t) = 1/(1+\exp(w (x-\phi)))^2$ and an ansatz function $\hat u(x,t) =  \operatorname{erfc}({w(x-\phi))}$; the accuracy was tested against simulations of the SPDE exhibiting the same order of magnitude in relative errors. The resulting equations for the evolution of the collective coordinates for these two test functions, however, are very cumbersome. \\ 
The method introduced here is not limited to travelling fronts. By choosing bell-shaped test functions rather than the front solutions explored in this work, one may study the front selection problem and propagation failure in stochastic systems supporting pulse solutions using collective coordinates as was done in the deterministic context \cite{CoxGottwald06}.\\  The symmetry perspective is not restricted to models with translational symmetry but can also be applied to pattern forming systems on the plane exhibiting Euclidean symmetry involving translations and rotations supporting spiral waves. This is relevant to stochastically perturbed excitable media. Spiral waves in excitable media can exhibit {\em meandering} of the spiral tip. Here the shape dynamics is periodic and the group dynamics evolves quasiperiodically. To apply the collective coordinate framework to this case is planned in further research.\\

The symmetry perspective has been successfully used in a numerical algorithm to simulate travelling waves in SPDEs \cite{LordThuemmler12}. Their algorithm, however, does not involve explicit expressions for the dynamics on the group, but the dynamics is determined numerically. To explicitly incorporate the dynamics of the collective coordinates within the freezing method introduced in \cite{BeynThuemmler04} might lead to a more efficient numerical method  to simulate SPDEs with symmetry.\\

Our work poses important theoretical questions. Whereas the symmetry perspective is well studied in the case of deterministic PDEs and conditions on its validity are well established (see for example \cite{GolubitskyStewart} and references therein), the extension into the stochastic realm of SPDEs with symmetry is only now developing 
with first promising rigorous results on front propagation with translational symmetry where a phase equation is sought to control the neutral direction \cite{HamsterHupkes17,HamsterHupkes18}. In the case of multiplicative noise for the Nagumo equation this leads to the same equations for the equilibrium shape and the phase we found in (\ref{e.bistable_mult_w})--(\ref{e.bistable_mult_c0}). 





\section*{Acknowledgments}
We thank Ben Goldys, Gabriel Lord, James Maclaurin and Gilles Vilmart for stimulating discussions. GAG acknowledges funding from the Australian Research Council, grant DP180101991.


\appendix
\section{Explicit formulae for the collective coordinate projections} 
\label{app.1}
We list here several integrals which appear in the evaluations of the projection when using the $\tanh$-ansatz function
\begin{align*}
\hat u(x,t) = \frac{1}{2}\left(1-\tanh(w(t)(x-\phi(t))) \right) .
\end{align*}
Using $u_x=-\pdif{u}{\phi}$ and $u_{xx}=\pdif{^2u}{\phi^2}$, we evaluate
\begin{align*}
	\pdif{u}{w}&=-\ha(x-\phi)\sechsq{w(x-\phi)},\\
	\pdif{u}{\phi}&=\ha w\sechsq{w(x-\phi)},\\
	\pdif{^2u}{w^2}&=(x-\phi)^2\sechsq{w(x-\phi)}\tanh\left(w(x-\phi)\right),\\
	\pdif{^2u}{\phi^2}&=w^2\sechsq{w (x - \phi)}\tanh\left({w (x - \phi)}\right),\\
	\pdif{^2u}{w\partial\phi}&=-w^2\sechsq{w(x-\phi)}\tanh{\left(w(x-\phi)\right)},\\
	u(1-u)&= \frac{1}{2w}\pdif{u}{\phi}=\frac{1}{4}\sechsq{w(x-\phi)} .
\end{align*}
The following integrals, which appear in the projections, can be analytically determined 
\begin{align*}
\langle \left(\pdif{u}{w}\right)^2\rangle & = -\frac{3}{2w} \langle \pdif{u}{w}\pdif{^2u}{w^2}\rangle= \frac{\pi^2-6}{36 w^3},\\
\langle \left(\pdif{u}{\phi}\right)^2\rangle & = 2 \langle u \left(\pdif{u}{\phi}\right)^2 \rangle = \frac{w}{3},\\
\langle \pdif{u}{w}\pdif{^2u}{\phi^2}\rangle & = -\langle \pdif{u}{\phi}\pdif{^2u}{w\partial \phi}\rangle=-\frac{1}{6},\\
\langle u\pdif{u}{w}\pdif{u}{\phi}\rangle & = \frac{1}{24 w},\\
\langle \pdif{u}{\phi}\pdif{u}{\phi}\rangle 
&= \langle \pdif{u}{\phi}\pdif{^2u}{\phi^2}\rangle 
=  \langle \pdif{u}{w}\pdif{^2u}{w \partial\phi}\rangle 
= \langle \pdif{u}{\phi}\pdif{^2u}{w^2}\rangle =0 .
\end{align*}

\section{Construction of $\bf{Q}$-Wiener noise}
\label{app.2}
Following closely \cite{Lord}, we briefly describe how to generate a $Q$-Wiener process, denoted  in this section by $\eta(t)$, with kernel of the covariance operator ${\mathcal{C}}(x,x^\prime)=\exp(-|x-x^\prime|/\ell)$ with finite trace. That is we need to find a complete orthonormal basis $\{\varphi_k\}_{k=1}^\infty$ as eigenfunctions of the covariance operator with $Q\varphi_k=\lambda_k\varphi_k$, to construct a $Q$-Wiener process as
\begin{align*}
\eta(x,t) = \sum_{k=1}^\infty \sqrt{\lambda_k}\varphi_k(x) B_k(t)
\end{align*} 
with independent one-dimensional Brownian motions $B_k(t)$.

Denote by $C=V^TV$ the $N\times N$ covariance matrix with $C=(c_{ij})$ with $c_{ij}=e^{|x_i-x_j|/\ell}$, approximating the covariance operator $Q$ and denote by $\deta$ the numerical approximation of the increments $d\eta(x,t)$ of the $Q$-Wiener process. We seek $\deta \sim\mathcal{N}(\mu,C)$ (in the main text we have $\mu=0$); then samples can be generated by
\begin{align*}
\deta = \mu+V^T\xi,
\end{align*}
where $\xi_j\sim\mathcal{N}(0,1)$. Suppose further that the spectral decomposition of $C$ is $C=U\Lambda U^T$, where $U$ is the orthonormal matrix with columns $\varphi_j$ being the eigenvectors of $C$, and $\Lambda$ is the diagonal matrix with eigenvalues $\lambda_j$. Let $V= U\Lambda^{\ha}$. Then 
\begin{align*}
\deta=\mu+\sum_{j=1}^N\sqrt{\lambda_j}\varphi_j\xi_j.
\end{align*}
A straightforward method to generate $\deta$ would be to find the eigenvectors and associated eigenvalues of the covariance matrix $C$, and truncate to $M$ eigenvectors and corresponding eigenvalues. However, this has complexity of $\Ord(N^3)$ to compute the eigen-decomposition. Computationally even more expensive is that at each time step a matrix multiplication of complexity $\Ord(NM)$ is required (see for example \cite{NumericalRecipes}). We note that as the correlation length $\ell$ decreases, the eigenvalues of $C$ decay more slowly so large values of $M$ are required to reliably approximate the $Q$-Wiener process.\\

We now describe a more efficient way to reduce the numerical complexity to $\Ord(N\log N)$ operations using the fast Fourier transform (FFT) by embedding the covariance matrix $C$ into a higher-dimensional circulant $2(N-1)\times 2(N-1)$ matrix. Let us define the following classes of matrices:
\begin{Definition}[Toeplitz matrix]
	A Toeplitz matrix is an $N\times N$ real valued matrix $C=(c_{ij})$ where $c_{ij}=c_{i-j}$ for some real numbers $c_{1-N},\ldots,c_{N-1}$.
\end{Definition}
\begin{Definition}[Circulant matrix]
	A Toeplitz matrix $C=(c_{ij})$ is circulant if $c_{ij}=c_{i-j}$ for $1\le j\le i$ and $c_{ij}=c_{i-j}=c_{1-j+N}$ for $i+1\le j\le N$.
\end{Definition}
Symmetric Toeplitz matrices have $c_{i-j}=c_{j-i}$, and symmetric circulant matrices have $c_{N-j}=c_j,\,j=1,2,\ldots,N-1$. Circulant matrices $C$ are of the following general form
\begin{align}
C=\begin{pmatrix}
c_0 & c_{N-1} & c_{N-2}& \cdots & c_2 & c_1\\
c_1 & c_0 & c_{N-1}  & \cdots & c_3 & c_2\\
\vdots & \ddots & \ddots & \ddots & \ddots & \vdots\\
c_{N-2} & c_{N-3} & \cdots & c_1 & c_0 & c_{N-1}\\
c_{N-1} & c_{N-2} & \cdots & c_{2} & c_1 & c_0
\end{pmatrix}.
\label{e.circulant}
\end{align}
One can diagonalize a real-valued circulant $N\times N$ matrix $C$ with first column $\mathbf{c}_1$ as $C=FDF^*$, where $F$ is the complex Fourier matrix (equivalent to the FFT) with entries $f_{\ell m}=\omega^{(\ell-1)(m-1)}/\sqrt{N}$ where $\omega=e^{-2\pi i/N}$, and $D$ is a diagonal with diagonal elements ${\mathbf{d}=\sqrt{N}F^*\mathbf{c}_1}$. For details the interested reader is referred to \cite{Lord}.  

In our application the covariance matrix for the $Q$-Wiener noise is $c_{ij}=e^{-|x_i-x_j|/\ell}=e^{-|i-j|\Delta x/\ell}$ and we have $c_{ij}=c_{i-j}=c_{j-i}$ and $c_{N-j}\neq c_j$. Therefore the covariance matrix $C$ is symmetric and Toeplitz, but it is not circulant. By embedding $C$ inside a larger circulant matrix $\tilde{C}\in\R^{2(N-1)\times2(N-1)}$, with
\[\tilde{C}=\begin{pmatrix}
C & B^T\\ B & D
\end{pmatrix}\]
where $B$ and $D$ are such that $\tilde{C}$ is circulant (i.e. of the form (\ref{e.circulant})). The larger circulant matrix $\tilde{C}$ has the decomposition $\tilde{C}=\tilde{F}\tilde{D}\tilde{F}^*$. $Q$-Wiener increments $\ddeta$ with mean 0 can be generated by $\ddeta=\tilde{F}\tilde{D}^{\ha}\xi$, where $\xi$ is a complex normal random variable, with $\re{\xi},\,\imag{\xi}\sim\mathcal{N}(0,1)$. Note that $\ddeta$ has independent identically distributed real and imaginary parts $\re{\ddeta},\,\imag{\ddeta}\sim\mathcal{N}(0,C)$, and so two independent samples can be generated simultaneously, which constitutes another computational advantage. 

Finally, we truncate $\ddeta$ to its first $N$ entries, and define $\deta=[\ddeta_1,\ddeta_2,\ldots,\ddeta_N]^T$. The covariance matrix of $\deta$ is therefore $C$, and we have the required $Q$-Wiener increment.



\end{document}